\documentclass[11pt,a4paper]{article}

\usepackage{graphicx}
\usepackage{epsfig}
\usepackage{amssymb}
\usepackage{amsmath}
\usepackage{float}
\usepackage{subfigure, amsmath,graphicx}
\usepackage{jheppub}

\newcommand{\bea}{\begin{eqnarray}}
\newcommand{\eea}{\end{eqnarray}}
\newcommand{\bean}{\begin{eqnarray*}}
\newcommand{\eean}{\end{eqnarray*}}
\newcommand{\nn}{\nonumber \\}

\def\O #1{\overline{#1}}

\def\W #1{\widetilde{#1}}
\def\WH #1{\widehat{#1}}

\def\braket#1{\left\langle #1 \right\rangle}
\def\bra#1{\left\langle #1\right|}
\def\ket#1{\left| #1\right\rangle}

\def\eref#1{(\ref{#1})}

\def\a{{\alpha}}

\def\la{\lambda}

\def\vev{\braket}

\def\Spaa{\vev}

\title{A note on  on-shell recursion relation of string amplitudes }

\author[a]{Yung-Yeh Chang,}
\author[b,c]{Bo Feng,}
\author[b,a]{Chih-Hao Fu,}
\author[a]{Jen-Chi Lee,}
\author[d]{Yihong Wang.}
\author[a]{Yi Yang}

\affiliation[a]{Department of Electrophysics, National Chiao Tung University\\
1001 University Street, Hsinchu, Taiwan, R.O.C.}
\affiliation[b]{Center of Mathematical Science, Zhejiang
University\\38 Zheda Road Hangzhou, 310027 P.R China}
\affiliation[c]{Zhejiang Institute of Modern Physics, Zhejiang
University\\38 Zheda Road Hangzhou, 310027 P.R China}
\affiliation[d]{Department of Physics and Astronomy, Stony Brook University\\
Stony Brook, NY 11794-3800, USA}
\emailAdd{cdshjtr.ep99g@nctu.edu.tw}
\emailAdd{b.feng@cms.zju.edu.cn}
\emailAdd{zhihao.fu@cms.zju.edu.cn}
\emailAdd{jcclee@cc.nctu.edu.tw}
\emailAdd{yihong.wang@stonybrook.edu}

\date{\today}
\abstract{ In the application of on-shell recursion relation to
string amplitudes, one challenge is the sum over infinite intermediate
on-shell string states. In this note, we show how to sum these
infinite states explicitly by including unphysical states to make
complete Fock  space.

}


\begin{document}
\maketitle

\section{Introduction}

Whilst its application requires merely the knowledge of analytic
structure of the scattering amplitude of interest, the on-shell
recursion  relation  (BCFW) \cite{{Britto:2004ap},Britto:2005fq} has
achieved tremendous success in calculations of scattering
amplitudes, a task would very often seem practically impossible
using conventional methods even when there are only a few of external
particles involving gluons or gravitons\footnote{A review of the
principles  of BCFW on-shell recursion relation
 as well as  its some applications
 can be found in \cite{Feng:2011gc}.}. In contrast to perturbative off-shell
 formulation, the on-shell recursion relation uses
fewer-point \textit{physical} amplitude as building blocks,
\begin{equation}
A(123\dots n)=\sum_{poles}A_{L}(\hat{1}2\dots,\hat{P}^{h})\,\frac{1}{P^{2}}\,
 A_{R}(-\hat{P}^{-h},\dots n),\label{eq:illustration}
\end{equation}
thereby avoiding large amount of unnecessary cancelation in intermediate
step of computations. An important point of Eq.
(\ref{eq:illustration}) is the sum over all possible physical poles
and allowed helicity configurations.  Generalization of on-shell
relation to string amplitudes was pioneered in
\cite{Boels:2008fc,Boels:2010bv} and  \cite{Cheung:2010vn} and
further elaborated in \cite{Fotopoulos:2010jz,Fotopoulos:2010cm, Fotopoulos:2010ay}.
Recent applications at $4$-point and to eikonal Regge limit 
can be found in \cite{Feng:2011qc} and \cite{Garousi:2010er} respectively. 
The validity of on-shell recursion relation in string theory context
was argued both from the better convergent UV behavior generically
observed in string amplitudes and from analyzing explicit
expressions of string amplitudes.

However, when applying on-shell recursion relation to string
amplitudes, we are facing the problem of summing over infinite
number of physical states in (\ref{eq:illustration}). Although it
could be done in principle, there is no efficient algorithm doing
so. For scattering amplitudes of tachyons, based on known analytic
expressions, it has been conjectured in \cite{Fotopoulos:2010jz}
that  amplitudes  can be effectively reduced to factorization of two
lower-point tachyon-like sub-amplitudes.

In this paper, we  provide an algorithm to do the sum over
infinity number of physical states in (\ref{eq:illustration}).
Applying our algorithm to tachyon amplitudes, we  see that the
sum over physical states at each mass level predicted by open string
theory does produce the conjectured scalar-behaved residue observed
in \cite{Cheung:2010vn}. In contrast with the experiences with
amplitude calculations in field theory, the key of our algorithm is
to enlarge the sum over intermediate physical states to over intermediate
complete Fock space states. The zero contributions of extra states are
guaranteed by no-ghost theorem (i.e., the Ward-like identity in
string theory)\footnote{We have summarized the no-ghost theorem in
Appendix \ref{ddf-ghost} for reference. }.

The structure of this paper is organized as the following: In
section \ref{BCFWreview}, we present a very brief review of BCFW
on-shell recursion relation of generic field theory amplitudes. In
section \ref{example1} we start with the familiar $4$-point
Veneziano amplitude as an example and demonstrate how the tachyonic
recursion relation can be understood from carrying out sum directly.
Section \ref{example2} consists of analysis on $5$-point string
amplitudes, in which case the pole structure becomes much more
complicated. A discussion on pole structure of generic $n$-point
amplitude is presented in section \ref{sec-n-pt}. In section
\ref{sec-vector} we consider higher-spin scatterings and demonstrate
that generically the mathematical connection between BCFW and
tachyonic recursion descriptions can be found in the generating function
for Stirling number of the first kind outlined in appendix \ref{app-mathid}, 
while the relation between
on-shell condition and decoupling of unphysical states is discussed
in appendix \ref{ddf-ghost}.

%%%%%%%%%%%%%%%%%%%%%%%%%%%%%%%%%%%%%%%%%%%%%%%%%%%%%%%
%%%%%%%%%%%%%%%%%%%%%%%%%%%%%%%%%%%%%%%%%%%%%%%%%%%%%%%
%%%%%%%%%%%%%%%%%%%%%%%%%%%%%%%%%%%%%%%%%%%%%%%%%%%%%%%
%%%%%%%%%%%%%%%%%%%%%%%%%%%%%%%%%%%%%%%%%%%%%%%%%%%%%%%
%%%%%%%%%%%%%%%%%%%%%%%%%%%%%%%%%%%%%%%%%%%%%%%%%%%%%%%
%%%%%%%%%%%%%%%%%%%%%%%%%%%%%%%%%%%%%%%%%%%%%%%%%%%%%%%

\section{A brief review of BCFW on-shell recursion relation}

\label{BCFWreview}

In this section we provide a short review of on-shell recursion
relation \cite{{Britto:2004ap},Britto:2005fq}. Derivation of BCFW
on-shell recursion relation starts from taking analytic continuation
of amplitudes. An amplitude can be regarded as function of complex
momenta defined by standard Feynman rules. When the momenta of a
pair of particle lines manually chosen are shifted in a complex
$q$-direction,
\bea \WH k_1(z)=k_1+z q,~~~~~\WH
k_n(z)=k_n-zq,~~~~\label{1n-deform}\eea
with $q^{2}=q\cdot k_{a}=q\cdot k_{n}=0$, the shifted amplitude
$A(z)$ defines a complex function. While the explicit analytic
structure of amplitude is determined by individual theory and does
not concern us here, $A(z)$ thus defined will contain simple poles
produced by propagators, which is the consequence of local
interaction and  the null condition of $q$. From Cauchy's Theorem,
integrating over a contour large enough to enclose all finite poles
yields
\begin{equation}
\oint dz\,\frac{A(z)}{z}=A(0)+\sum_{poles\,\alpha}\,
Res_{z=z_{\alpha}},\label{eq:cauchy integral}
\end{equation}
where an unshifted amplitude $A(0)$ contributes as residue at $z=0$
and residues from other finite poles assume the form as
cut-amplitudes,
$Res_{z_{\alpha}}=-A(z_{\alpha})\frac{1}{P^{2}}A_{R}(z_{\alpha})$.
In various theories shifted amplitudes posses convergent large-$z$
asymptotic behavior and the integral (\ref{eq:cauchy integral})
vanish, we are then entitled to write down the BCFW recursion
relation\footnote{We have assumed the boundary contribution to be
zero. If it is no zero, we need to modify recursion relation, see
\cite{Feng:2009ei}.}
\bea A_n =\sum_{poles}\sum_{
{\tiny
\begin{array}{c}
physical \\
states
\end{array}
 }
  } A_L(..., P(z_\a)) {2\over
P^2+M^2} A_R( -P(z_\a),...),~~~\label{BCFW-form}\eea
where the first sum is over all finite simple poles $z_\a$ of $z$,
and the second sum is over all physical states at the given simple
pole $z_a$.

%%%%%%%%%%%%%%%%%%%%%%%%%%%%%%%%%%%%%%%%%%%%
%%%%%%%%%%%%%%%%%%%%%%%%%%%%%%%%%%%%%%%%%%%%%
%%%%%%%%%%%%%%%%%%%%%%%%%%%%%%%%%%%%%%%%%%%%%%%%%%%%%%%
%%%%%%%%%%%%%%%%%%%%%%%%%%%%%%%%%%%%%%%%%%%%%%%%%%%%%%%
%%%%%%%%%%%%%%%%%%%%%%%%%%%%%%%%%%%%%%%%%%%%%%%%%%%%%%%
%%%%%%%%%%%%%%%%%%%%%%%%%%%%%%%%%%%%%%%%%%%%%%%%%%%%%%%
%%%%%%%%%%%%%%%%%%%%%%%%%%%%%%%%%%%%%%%%%%%%%%%%%%%%%%%
%%%%%%%%%%%%%%%%%%%%%%%%%%%%%%%%%%%%%%%%%%%%%%%%%%%%%%%

%%%%%%%%%%%%%%%%%%%%%%%%%%%%%%%%%
\section{Example I: BCFW of $4$-tachyon amplitude in bosonic open string theory}
%%%%%%%%%%%%%%%%%%%%%%%%%%%%%%%%%
\label{example1}
As was demonstrated in the previous section, a key feature making BCFW
on-shell recursion relation possible is that in perturbative field theory, at tree-level
amplitude can often be determined entirely from its poles and related residues.
The locations of poles are determined by propagators while the residues, by
factorization properties. Same analytic structure holds for string theory, with one
complication: there is an infinite number of poles and related residues. As an consequence,
there are several expressions for amplitudes, for example, the    Veneziano formula assumes
the form of a worldsheet integral, making the  pole structure obscured. In
\cite{Cheung:2010vn} through binomial expansions of these integral formulas, the pole structure
can be made manifest. In this section, we will use four-point tachyon amplitude as an example
to demonstrate our idea and method.

%%%%%%%%%%%%%%%%%%%%%%%%%%
\subsection{Pole structure extraction  }
%%%%%%%%%%%%%%%%%%%%%%%%%%

Consider the four tachyon scattering amplitude in
bosonic open string theory, given by Koba-Nielson formula as
\begin{equation}
A(1234)=\int_0^1 dz_{2}\, (1-z_{2})^{k_{3}\cdot k_{2}}\,
z_{2}^{k_{2}\cdot k_{1}},~~\label{eq:4pt}
\end{equation}
where we have used the conformal symmetry to fix $z_1=0$, $z_3=1$
and $z_4=+\infty$. For arbitrary complex power $w$ we have following
binomial expansion
\bea (x-y)^w=\sum_{a=0}^\infty \left(\begin{array}{c} w\\ a
\end{array}\right) x^{w-a} y^a~~\label{binomial}\eea
where coefficient $\left(\begin{array}{c} w\\ a
\end{array}\right)$ is defined as
\bea \left(\begin{array}{c} w\\ a
\end{array}\right)= { w(w-1)(w-2)...(w-a+1)\over a!}\eea
Applying \eref{binomial} to $(1-z_{2})^{k_{3}\cdot k_{2}}$ and
collecting relative terms we have
\begin{equation}
A(1234)=\sum_{a=0}^\infty \left(\begin{array}{c}
k_{3}\cdot k_{2}\\
a
\end{array}\right)(-)^{a}\int dz_{2}\, z_{2}^{\frac{1}{2}(k_{1}+k_{2})^{2}+a-2}
\end{equation}
where we have used the mass-shell condition for tachyon that
$k_1^2=k_2^2=-M^2=+2$\footnote{We have used the convention
$\a'=1/2$, so the mass of bosonic open string state is
$M^2=-2+2\sum_{n=1}^\infty \a_{-n}\cdot \a_n$}. The worldsheet
integration can be explicitly carried out, producing an $s$-channel
propagator\footnote{In this expansion, only $s$-channel is manifest. However,
by string duality, $t$-channel is also contained. }. %
%\begin{equation}
%\int_{0}^{1}dz_{2}\,
%z_{2}^{\frac{1}{2}(k_{1}+k_{2})^{2}+a-2}=\frac{1}{\frac{1}{2}(k_{1}+k_{2})^{2}+a-1}.
%~~\label{eq:spole}
%\end{equation}
%
Inserting it back, we obtain
\begin{equation}
A(1234)=\sum_{a=0}^\infty \left(\begin{array}{c}
k_{3}\cdot k_{2}\\
a
\end{array}\right)(-)^{a}\frac{2}{(k_{1}+k_{2})^{2}+2(a-1)}~~\label{4-pole-form}
\end{equation}

~\\
%%%%%%%%%%%%%%%%%%%%%%%%%%%%%%%%
%%%%%%%%%%%%%%%%%%%%%%%%%%%%%%%
%%%%%%%%%%%%%%%%%%%%%%%%%%%%%%%%
%%%%%%%%%%%%%%%%%%%%%%%%%%%%%%%%
%%%%%%%%%%%%%%%%%%%%%%%%%%%%%%%
%%%%%%%%%%%%%%%%%%%%%%%%%%%%%%%%
%%%%%%%%%%%%%%%%%%%%%%%%%%%%%%%%
%%%%%%%%%%%%%%%%%%%%%%%%%%%%%%%
%%%%%%%%%%%%%%%%%%%%%%%%%%%%%%%%
\subsection{Interpreting pole expansion formula from BCFW perspective}
\label{section-int-pole}
%%%%%%%%%%%%%%%%%%%%%%%%%%%%%%

Having derived an explicit analytic expression \eref{4-pole-form}
for tree-level four tachyon scattering amplitude, it is then
interesting to see if the result can be understood in the language
of BCFW on-shell recursion relation. We choose the shifted pair to
be $(1,4)$ to be consistent with the manifest $s$-channel expansion.
Assuming there is no boundary
contribution for on-shell recursion relation, equation
 \eref{4-pole-form} should be given by on-shell  recursion relation \eref{BCFW-form}:
\bea A_n =\sum_{poles}\sum_{physical} A_L(..., P(z_\a)) {2\over
P^2+M^2} A_R( -P(z_\a),...)~~~\label{BCFW-form2}\eea
In denominator we see infinitely many single poles occurs at
\bea z_a= {(k_1+k_2)^2+2(a-1)\over -2 q\cdot
(k_1+k_2)},~~~~a=0,1,\dots .~~~~\label{mass-level}\eea
where $P=k_1+k_2$ and  the mass square  $M_a^2=2(a-1)$ for every
integer $a$  is precisely the mass spectrum prescribed by bosonic
open string theory. In addition, matching residues of
\eref{4-pole-form} with \eref{BCFW-form2} indicates that, at each
level $a$, there should be a number of physical states, collectively
yielding
\bea \sum_{states~~h} A_L(1,2,P_a^h(z_a))A_R(-P_a^{\W
h}(z_a),3,4)=(-1)^a\left(\begin{array}{c}
k_{3}\cdot k_{2}\\
a
\end{array}\right).~~~~\label{4point-heli-sum}
\eea
Thus to understand \eref{4-pole-form} from BCFW recursion relation
\eref{BCFW-form}, we need to be able to interpret the scalar-behaved
residue
\eref{4point-heli-sum} as sum over
physical states at each fixed level $a$.

%%%%%%%%%%%%%%%%%%%
\subsection{ Summing over physical states}
%%%%%%%%%%%%%%%%%%%%

Before undertaking a state-by-state calculation of residues over
bosonic string spectrum, let us make a slight detour and consider
how the analytic structure featuring intermediate states fits into
the picture of BCFW on-shell recursion relation in quantum field
theory. Although in Feynman rules scalar, fermion and gauge boson
each are assigned with a propagator in distinct representations, we
note that the propagator appearing in BCFW recursion relation
\eref{BCFW-form2} is always scalar-like. The reason is following.
For example, if the intermediate particles are massless fermions,
BCFW recursion relation reads
\bea A\sim \sum_{h=\pm} A_L(\sigma_L, P^{h}) A_R (-P^{-h}, \sigma_R).
\eea
We can rewrite the on-shell sub-amplitude $A_L(\sigma_L, P^{h})=
\sum_{a=1,2}\W A_L(\sigma_L, P^{h})^{a} u^h(P)_a$, i.e., we have
decomposed the on-shell amplitude into two parts: wave function for
external on-shell particle $P$ and the rest. Similar decomposition
can be done for $A_R (-P^{-h}, \sigma_R)$. Thus the sum over
physical states becomes
\bea A\sim \W A_L(\sigma_L, P^{h}) \left(\sum_h u^s(P) \O
u^s(p)\right) \W A_R (-P^{-h}, \sigma_R)\sim \W A_L(\sigma_L, P^{h})
\left(\gamma\cdot P\right) \W A_R (-P^{-h}, \sigma_R)\eea
where in the middle, $\gamma\cdot P$ is exactly the factor needed to
translate  scalar propagator into the familiar fermion propagator.

A similar mechanism supports the translation from scalar propagator
into gauge boson propagator when summed over physical states, but
with some subtleties. The sum over two transverse physical states
for gauge boson is $(\epsilon_\mu^+ \epsilon_\nu^- +\epsilon_\mu^-
\epsilon_\nu^+)$ while the familiar Feynman gauge uses $g_{\mu\nu}$.
In fact, in 4-dimensions we need four polarization vectors, and
\bea g_{\mu\nu}= \epsilon_\mu^+ \epsilon_\nu^- +\epsilon_\mu^-
\epsilon_\nu^+ +\epsilon_\mu^L \epsilon_\nu^T+\epsilon_\mu^T
\epsilon_\nu^L ~~~\label{vector}\eea
where $\epsilon_\mu^L$ and $\epsilon_\mu^T$ are longitude and
time-like polarization vector \cite{Feng:2011twa}. The reason that
these two sums (Namely a
 summation over two physical states and another over all four states) give
same answer depends crucially on Ward Identity of gauge theory,
i.e., if all $(n-1)$ particles are physical polarized while the
$n$-th particle is longitude (i.e., proportional to $k_\mu$), the
amplitude is zero. Thus we have
\bea & & \sum_{all~states} A_L(\sigma_L, P^h)  A_R(-P^{\W h},
\sigma_R)  \sim  \W A_L^\mu(\sigma_L, P) g_{\mu \nu} \W A_R^\nu(-P,
\sigma_R)\nn
&\sim &  \W A_L^\mu(\sigma_L, P) \left( \epsilon_\mu^+
\epsilon_\nu^- +\epsilon_\mu^- \epsilon_\nu^+ +\epsilon_\mu^L
\epsilon_\nu^T+\epsilon_\mu^T \epsilon_\nu^L\right)\W A_R^\nu (-P,
\sigma_R)  \sim   \W A_L^\mu(\sigma_L, P) \left( \epsilon_\mu^+
\epsilon_\nu^- +\epsilon_\mu^- \epsilon_\nu^+ \right)\W A_R^\nu (-P,
\sigma_R)\nn
& = & \sum_{physical~states} A_L(\sigma_L, P^h)  A_R(-P^{- h},
\sigma_R)\eea

Having understood the effect of summing over physical states from
quantum field theory, let us return to the problem of interpreting
scalar-behaved residue \eref{4-pole-form} as sum over physical
states. In old covariant quantization framework, the Fock space in
bosonic open string theory is constructed by linear combinations of
states  obtained from acting creation modes successively on ground
state
\begin{equation}
\alpha_{-n_{1}}^{\mu_{1}}\alpha_{-n_{2}}^{\mu_{2}}\dots
\alpha_{-n_{n}}^{\mu_{n}}\ket{0;k}.~~~\label{eq:fockstate1}
\end{equation}
Generically, a Fock state can carry $N_{\mu,1}$-multiple of
$\alpha_{-1}^{\mu}$ mode
operators\footnote{It should be emphasized that $\a_{-1}^\mu$ and
$\a_{-1}^\nu$ should be considered as different operators when
$\mu\neq \nu$.} and $N_{\mu,2}$-multiple of $\alpha_{-2}^{\mu}$ mode
and so on. In the following discussions we use the set of
numbers $\{N_{\mu,n}\}$ as  label of normalized Fock state
\begin{equation}
\begin{array}{cccccc}
\ket{\{ N_{\mu,n}\},\, k}= & \left[ \prod_{\mu=0}^{D-1}
\prod_{n=1}^\infty { (\a^\mu_{-n})^{N_{\mu,n}}\over
\sqrt{n^{N_{\mu,n}} N_{\mu,n}!}}\right]
%\prod\frac{\left(\alpha_{-1}^{\mu}\right)^{n_{\mu,1}}}{\sqrt{n_{\mu,1}!\,1^{_{\mu,1}}}}&
%\frac{\left(\alpha_{-2}^{\mu}\right)^{n_{\mu,2}}}{\sqrt{n_{\mu,2}!\,2^{n_{\mu,2}}}}
%&\dots&\frac{\left(\alpha_{-t}^{\mu}\right)^{n_{\mu,t}}}{\sqrt{n_{\mu,t}!\,
%t^{N_{\mu,t}}}} & \,
\ket{0,\, k}.\end{array}~~~\label{eq:fockstate}
\end{equation}
Physical states however, in addition must satisfy Virasoro constraints
% \bea
$(L_0-1)\ket{\phi}=0$, %~~~~
$L_{m>0}\ket{\phi}=0$
%\eea
%
and constitute only a subset in Fock space.
An immediate consequence is that physical states
are automatically on the mass-shell, $-k^2=M^2=2(N-1)$, where
$N$ is the level
\bea N=\sum_{\mu=0}^{D-1} \sum_{n=1}^\infty n
N_{\mu,n}~.~~~~\label{Level-N}\eea
%
%In the language of vertex operator, the physical vertex operator
%must be primary operator with weight $h=1$.
Note however, for a generic Fock state its center-of-mass momentum
$k^{\mu}$  and modes $\left\{ N_{\mu,n}\right\}$
are considered as independent degrees of freedom and does not
{\it a priori} %have to
satisfy mass-shell condition, and yet in a BCFW on-shell recursion
relation, Fock states that happen to be the on mass-shell are picked
out because as we have seen from \eref{mass-level} that only these
states contribute to residues.

Now we come to our central point. The prescription given by BCFW
on-shell recursion relation is to sum over physical states
satisfying on-shell condition plus  remaining Virasoro constraints
$L_{m>0}\ket{\phi}=0$. However, a rather technical difficulty
carrying out above prescription in string theory
 is that it requires the knowledge of physical polarization tensor
at arbitrarily high mass level $N$, which is very hard to write down
explicitly. To bypass the problem, inspired by the observation
given in \cite{Feng:2011twa} for gauge theory \eref{vector}, we can
enlarge the sum over physical states to all states in Fock space
satisfying on-shell condition. The fact that these two sums are same
is guaranteed by the famous ``No-Ghost Theorem''\footnote{We have
collected some facts of ``No-Ghost Theorem'' in appendix
\ref{ddf-ghost}.}. With this understanding, we can write
\bea A_n & = & \sum_{poles}\sum_{physical} A_L(..., P(z_\a)) {2\over
P^2+M^2} A_R( -P(z_\a),...)\nn
& = & \sum_{poles}\sum_{Fock} A_L(..., P(z_\a)) {2\over P^2+M^2}
A_R( -P(z_\a),...)\nn
& = & \sum_{poles}\sum_{Fock} (\W A_L(..., P(z_\a)) \cdot \xi_P
{2\over P^2+M^2} \W A_R( -P(z_\a),...)\cdot
\xi^*(P)~~~\label{BCFW-form-1}\eea
where at the last step we have stripped away the polarization tensor
of intermediate state $P$ from on-shell amplitude. Since the sum is taken over
whole Fock space, we are free to choose any convenient basis, for example,
the one given in \eref{eq:fockstate}, to perform the sum. Thus if we take
pair $(1,n)$ to conduct BCFW-deformation and sum over the polarization
tensor of intermediate state, BCFW on-shell relation of a string amplitude
reads
\bea A_n & = & \sum_{i=2}^{n-2}\sum_{N=0}^{+\infty}
\sum_{\{N_{\mu,n}\}}\Spaa{\phi_1(\WH k_1)|
V_2(k_2)...V_i(k_i)|\{N_{\mu,n}\},\WH P} {2{\cal
T}_{\{N_{\mu,n}\}}\over (\sum_{t=1}^i k_i)^2+2(N-1)} \nn & &
\Spaa{\{N_{\mu,n}\},\WH
P|V_{i+1}(k_{i+1})...V_{n-1}(k_{n-1})|\phi_n(\WH k_n)}
~~~~\label{String-BCFW}\eea
In this formula, the first sum is over the splitting of particles
into left and right handed sides while the second sum is over poles
fixed by the mass level $N$. The third sum is over all allowed
choice of the set $\{N_{\mu,n}\}$ as long as they satisfy
\eref{Level-N}. The tensor structure ${\cal T}_{\{N_{\mu,n}\}}$ is
determined by the set $\{N_{\mu,n}\}$. To demonstrate the rule for
the tensor structure, we list the tensor structure for first three
levels:
\begin{itemize}

\item {\bf Level $N=0$:} For the first level, all $N_{\mu,n}=0$ so
we have ${\cal T}=1$.

\item {\bf Level $N=1$:} The choice is $N_{\mu,1}=1$ for
$\mu=0,1,...,D-1$, thus we have ${\cal T}=g_{\mu\nu}$, i.e., we have
\bea \Spaa{\phi_1|... V_i~ {\a_{-1}^\mu}|0;P} { 2g_{\mu\nu}\over
P^2+2(N-1)} \Spaa{0;P|
{\a_{+1}^\nu}~V_{i+1}...|\phi_n}~~~\label{4Level-1}\eea
where when we conjugate $\ket{{\a_{-1}^\mu}|0;P}$ we get $\bra{0;P|
{\a_{+1}^\nu}}$

\item {\bf Level $N=2$:} There are several choices and the structure is
given by
\bea & & \sum_{\mu,\nu=0}^{D-1}\Spaa{\phi_1|... V_i~
{\a_{-2}^\mu\over \sqrt{2}}|0;P} {2 g_{\mu\nu}\over P^2+2(N-1)}
\Spaa{0;P| {\a_{+2}^\nu\over \sqrt{2}}~V_{i+1}...|\phi_n}\nn
& + & \sum_{0\leq \mu_1 <\mu_2\leq D-1}\sum_{0\leq \nu_1 <\nu_2\leq
D-1}\Spaa{\phi_1|... V_i~ \a_{-1}^{\mu_1}\a_{-1}^{\mu_2}|0;P} {2
g_{\mu_1\nu_1}g_{\mu_2\nu_2}\over P^2+2(N-1)} \Spaa{0;P|
\a_{+1}^{\nu_2}\a_{+1}^{\nu_1}~V_{i+1}...|\phi_n}\nn
& + & \sum_{\mu,\nu=0}^{D-1}\Spaa{\phi_1|... V_i~
{(\a_{-1}^\mu)^2\over \sqrt{2}}|0;P} { 2 (g_{\mu\nu})^2\over
P^2+2(N-1)} \Spaa{0;P| {(\a_{+1}^\nu)^2\over
\sqrt{2}}~V_{i+1}...|\phi_n}~~~\label{4Level-2}\eea
where at the second line, to avoid repetition, we must have the
ordering $0\leq \mu_1 <\mu_2\leq D-1$.

\item {\bf Level $N=3$:} There are several choices which are
given respectively by

\bean T_1&= & \sum_{\mu,\nu=0}^{D-1}\Spaa{\phi_1|... V_i~
{\a_{-3}^\mu\over \sqrt{3}}|0;P} {2 g_{\mu\nu}\over P^2+2(N-1)}
\Spaa{0;P| {\a_{+3}^\nu\over \sqrt{3}}~V_{i+1}...|\phi_n}\eean
\bean T_2&= &
\sum_{\mu_{1},\mu_2,\nu_{1},\nu_2=0}^{D-1}\Spaa{\phi_1|... V_i~
{\a_{-2}^{\mu_1}\over \sqrt{2}}\a_{-1}^{\mu_2}|0;P} {2
g_{\mu_1\nu_1}g_{\mu_2\nu_2}\over P^2+2(N-1)} \Spaa{0;P|
\a_{+1}^{\nu_2}{\a_{+2}^{\nu_1}\over
\sqrt{2}}~V_{i+1}...|\phi_n}\eean
\bean T_3& = & \sum_{0\leq \mu_1 <\mu_2<\mu_3\leq D-1}\sum_{0\leq
\nu_1 <\nu_2<\nu_3\leq D-1}\Spaa{\phi_1|... V_i~
\a_{-1}^{\mu_1}\a_{-1}^{\mu_2}\a_{-1}^{\mu_3}|0;P}\nn & &  {2
g_{\mu_1\nu_1}g_{\mu_2\nu_2}g_{\mu_3\nu_3}\over P^2+2(N-1)}
\Spaa{0;P|
\a_{+1}^{\nu_3}\a_{+1}^{\nu_2}\a_{+1}^{\nu_1}~V_{i+1}...|\phi_n}\eean
\bean T_4 & = &
\sum_{\mu_{1},\mu_2,\nu_{1},\nu_2=0}^{D-1}\Spaa{\phi_1|... V_i~
{(\a_{-1}^{\mu_1})^2\over \sqrt{2}} (a_{-1})^{\mu_2}|0;P} { 2
(g_{\mu_1\nu_1})^2 g_{\mu_2\nu_2}\over P^2+2(N-1)} \Spaa{0;P|
(\a_{+1})^{\nu_2}{(\a_{+1}^{\nu_1})^2\over
\sqrt{2}}~V_{i+1}...|\phi_n}\eean
\bean T_5 & = & \sum_{\mu,\nu=0}^{D-1}\Spaa{\phi_1|... V_i~
{(\a_{-1}^{\mu})^3\over \sqrt{3!}} |0;P} { 2 (g_{\mu\nu})^3 \over
P^2+2(N-1)} \Spaa{0;P| {(\a_{+1}^{\nu})^3\over
\sqrt{3!}}~V_{i+1}...|\phi_n}\eean
So we have
\bea N=3:~~~T_1+T_2+T_3+T_4+T_5~~~~~~~\label{4Level-3}\eea

\end{itemize}

These examples demonstrate  the general pattern of  tensor
structures. However, because when we have several oscillators with
same $n$, there are freedoms with the choice of $\mu$,  we need to
distinguish if these $\mu$ are same or different from each other.
This makes the tensor structure a little bit of complicated. This
complication can be simplified further. For example, at the level
$N=2$, we have
\bea & & \sum_{0\leq \mu_1\leq \mu_2\leq D-1}\sum_{0\leq \nu_1\leq
\nu_2\leq D-1} \a_{-1}^{\mu_1}\a_{-1}^{\mu_2} g_{\mu_1 \nu_1}
g_{\mu_2 \nu_2}\a_{+1}^{\nu_2} \a_{+1}^{\nu_1}  =   {1\over
2}\sum_{\mu_1\neq \mu_2=0}^{D-1}\sum_{\nu_1\neq \nu_2=0}^{D-1}
\a_{-1}^{\mu_1}\a_{-1}^{\mu_2} g_{\mu_1 \nu_1}g_{\mu_2 \nu_2}
\a_{+1}^{\nu_2} \a_{+1}^{\nu_1}\nn
& = & \sum_{\mu_1\neq \mu_2=0}^{D-1} \sum_{\nu_1\neq \nu_2=0}^{D-1}
{\a_{-1}^{\mu_1}\a_{-1}^{\mu_2}\over \sqrt{2}} g_{\mu_1
\nu_1}g_{\mu_2 \nu_2} {\a_{+1}^{\nu_2} \a_{+1}^{\nu_1}\over
\sqrt{2}}\eea
With this rewriting, the second and third line of \eref{4Level-2}
can be combined to
\bea \sum_{\mu_1,\mu_2,\nu_1,\nu_2=0}^{D-1}\Spaa{\phi_1|... V_i~
{\a_{-1}^{\mu_1}\a_{-1}^{\mu_2}\over \sqrt{2}}|0;P} { 2
g_{\mu_1\nu_1}g_{\mu_2\nu_2}\over P^2+2(N-1)} \Spaa{0;P|
{\a_{+1}^{\nu_1}\a_{+1}^{\nu_2}\over
\sqrt{2}}~V_{i+1}...|\phi_n}\eea
Similar argument can show that the sum $T_3, T_4, T_5$ of
\eref{4Level-3} gives
\bean T_3+T_4+T_5& = & \sum_{ \mu_i, \nu_i=0}^{D-1}\Spaa{\phi_1|...
V_i~ {\a_{-1}^{\mu_1}\a_{-1}^{\mu_2}\a_{-1}^{\mu_3}\over
\sqrt{3!}}|0;P}{2 g_{\mu_1\nu_1}g_{\mu_2\nu_2}g_{\mu_3\nu_3}\over
P^2+2(N-1)} \Spaa{0;P|
{\a_{+1}^{\nu_3}\a_{+1}^{\nu_2}\a_{+1}^{\nu_1}\over
\sqrt{3!}}~V_{i+1}...|\phi_n}\eean
It is easy to see that when multiple operators of the same mode $n$  
are present in the Fock state, each may or may not be carrying the same 
Lorentz index $0$ , or $1$ , or $\dots$ , or $D-1$,
the general pattern is given by the expansion
$(a_0+a_1+...+a_{D-1})^{N_{n}}/N_{n}!$ where $a_i= \a_{-n}^{i} g_{ii}
\a_{+n}^{i}$. The coefficient of term $(\alpha_{-n}^{0})^{n_0} (\alpha_{-n}^{1})^{n_1}...
(\alpha_{-n}^{D-1})^{n_{D-1}}$ in the Fock state is given by the 
coefficient of term $a_0^{n_0} a_1^{n_1}...
a_{D-1}^{n_{D-1}}$ with $N_{n} =\sum_{i=0}^{D-1} n_i$ in the expansion,
which reads
\bea {1\over N!} C^{N}_{n_0}
C^{N-n_0}_{n_1} C^{N-n_0-n_1}_{n_2}...C^{n_{D-1}}_{n_{D-1}}={1\over
N!} {N!\over  \prod_{i=0}^{D-1} (n_i)!}\eea
thus we can drop the $\mu_{1}<\mu_{2}< \dots $ arrangement and
% have ${a_i\over (n_i)!}={\a_{-t}^{i}\over \sqrt{(n_i)!}}
% g_{ii} {\a_{+t}^{i}\over \sqrt{(n_i)!}}$.
%
% With above simplification, we can 
rewrite the sum in
\eref{String-BCFW} as
\bea & & \sum_{\{N_{\mu,n}\} } \ket{\{N_{\mu,n}\};\WH P} {\cal
T}_{\{N_{\mu,n}\}} \bra{\{N_{\mu,n}\};\WH P}\nn
& = & \sum_{\sum_n nN_{n}=N } \left\{\prod_{n=1}^\infty {
(\a_{-n}^{{\mu_{N_{n},1}}} \a_{-n}^{{\mu_{N_{n},2}}}...
\a_{-n}^{{\mu_{N_{n},N_{n}}}})\over \sqrt{N_{n}!
n^{N_{n}}}}\right\}\ket{0;\WH P} \prod_{n=1}^\infty
(g_{{\mu_{N_{n},1}}{\nu_{N_{n},1}}}
g_{{\mu_{N_{n},2}}{\nu_{N_{n},2}}}...
g_{{\mu_{N_{n},N_{n}}}{\nu_{N_{n},N_{n}}}})\nn
& & \bra{0;\WH P} \left\{\prod_{n=1}^\infty {
(\a_{+n}^{{\nu_{N_{n},1}}} \a_{+n}^{{\nu_{N_{n},2}}}...
\a_{+n}^{{\nu_{N_{n},N_{n}}}})\over \sqrt{N_{n}!
n^{N_{n}}}}\right\}~~~\label{tensor-structure}\eea
Having the simplified version \eref{tensor-structure}, we can give
following explicit calculations.

%%%%%%%%%%%%%%%%%%%%%%%%
\subsubsection{ Explicit calculation}
%%%%%%%%%%%%%%%%%%%%%%%

 Recalling the vertex of tachyon
\bea V_0(k,z)= : e^{ik\cdot X(z)} := Z_0 W_0~,
~~~~\label{tachyon}\eea
where
\bea Z_0 & = & e^{i k\cdot x+k\cdot p\ln z}= e^{ikx} z^{k\cdot p+1}=
z^{k\cdot p-1} e^{ik\cdot x}\eea
and
\bea W_0 & = & e^{ \sum_{n=1}^\infty {z^n\over n} k\cdot\a_{-n}} e^{
-\sum_{n=1}^\infty {z^{-n}\over n} k\cdot\a_{n}}~,\eea
it is easy to calculate the left three-point amplitude
\bea \Spaa{0;-k_1| V_0(k_2, z)|\{N_{\mu,n}\}; P} & =
&\delta(k_1+k_2+P) \prod_{\mu=0}^{D-1}\prod_{m=1}^\infty { (
-k_2^\mu)^{N_{\mu,m}}\over \sqrt{
m^{N_{\mu,m}}N_{\mu,m}!}}~~~\label{12P}\eea
where $N$ is the level defined in \eref{Level-N} and the right
three-point amplitude
\bea \Spaa{\{N_{\mu,n}\}; P| V_0(k_3,z)| 0;k_4}=
\delta(P-k_3-k_4)\prod_{\mu=0}^{D-1} \prod_{m=1}^\infty { (
k_3^\mu)^{N_{\mu,m}}\over \sqrt{ {N_{\mu,m}}!
m^{N_{\mu,m}}}}~~~\label{P34}\eea
Using \eref{12P} and \eref{P34} it is easy to calculate first few
mass levels. In fact, the same calculation has been done in our
simplification leading to the simplified tensor structure
\eref{tensor-structure}. Thus we have when $N=0$, it is $1$, while
when $N=1$ it is $(-k_2\cdot k_3)$. Finally when $N=2$ it is ${
(k_2\cdot k_3)(k_2\cdot k_3-1)\over 2}$. They do satisfy
\eref{4point-heli-sum} for $N=0,1,2$.

For general level $N$, from \eref{tensor-structure}, \eref{12P} and
\eref{P34} we find
\bea I_N=\sum_{\sum nN_{n}=N} \prod {(-k_2\cdot k_3)^{N_n}\over
N_{n}! n^{N_{n}}}\eea
Let us define
\bea N=\sum_{n=1}^\infty n N_{n},~~~~~~J=\sum_{n=1}^\infty N_n\eea
with obviously that $J\leq a$, then using the definition
\eref{Stirling-1-exp} of Stirling number of the first kind, $I_N$
can be rewritten as
\bea I_N=(-)^N\sum_{J=1}^N { S(N,J)\over N!} (k_2\cdot k_3)^J= (-)^N
\left( \begin{array}{c} k_2\cdot k_3 \\  N\end{array}\right) ~~~\label{4-pt-res}\eea
where we have used the formula \eref{Stirling-1-def}\footnote{Since
$S(N,0)=0$ when $N>0$, we can extend the sum over $J$ from region
$[1,N]$ to region $[0,N]$.}. This is exactly the result
\eref{4point-heli-sum} we try to prove.

%%%%%%%%%%%%%%%%%%%%%%%%%%%%%
%%%%%%%%%%%%%%%%%%%%%%%%%%%%%%
%%%%%%%%%%%%%%%%%%%%%%%%%%%%%%%
%%%%%%%%%%%%%%%%%%%%%%%%%%%%%%%
%%%%%%%%%%%%%%%%%%%%%%%%%%%%%%%%%
%%%%%%%%%%%%%%%%%%%%%%%%%%%%%%%%%

%%%%%%%%%%%%%%%%%%%%%%%%%%%%%
\section{Example II: BCFW of $5$-tachyon amplitude in bosonic open string theory}
\label{example2}
%%%%%%%%%%%%%%%%%%%%%%%%%%%%%

Having shown that a 4-point Veneziano amplitude can be indeed
described by BCFW on-shell recursion relation, let us  consider the
5-tachyon scattering amplitude, which contains slightly richer
analytic structure because unlike  4-point amplitude with only pole
$s_{12}$, there are two types of poles from $s_{12}, s_{123}$ for
deformation (1,5). Multiple pole structure is seen  for general
amplitudes, we need to study this simplest nontrivial example.

%%%%%%%%%%%%%%%%%%%%%
\subsection{Pole expansion}
%%%%%%%%%%%%%%%%%%%%

%{\bf Expanded Form:}
The Koba-Nielson formula for $5$-point tachyon
amplitude is given by
\begin{equation}
A(12345)=\int_{0}^{1}dz_{3}\int_{0}^{z_{3}}dz_{2}\,(1-z_{3})^{k_{4}\cdot
k_{3}}(1-z_{2})^{k_{4}\cdot k_{2}}(z_{3}-z_{2})^{k_{3}\cdot
k_{2}}z_{2}^{k_{2}\cdot k_{1}}z_{3}^{k_{3}\cdot
k_{1}}.~~~\label{eq:nielson}
\end{equation}
where we have fixed $z_1=0, z_4=1, z_5=\infty$. Unlike in quantum
field theory, where analytic behavior of an amplitude is transparent
from Feynman rules, kinematic dependence in Koba-Nielson's
formulation were implicitly introduced through exponents of
worldsheet integration variables, making it less easier to locate
 poles. However as we have seen in the previous section,
worldsheet integrals can be explicitly carried out after binomial
expansions. Expanding $(z_{3}-z_{2})^{k_{3}\cdot k_{2}}$ with
respect to $z_{2}$, which is the variable that assumes smaller value
(than $z_{3}$), and expand similarly $(1-z_{2})^{k_{4}\cdot k_{2}}$
and $(1-z_{3})^{k_{4}\cdot k_{3}}$ we have
\bea (1-z_{2})^{k_{4}\cdot k_{2}} & = &
\sum_{a=0}^{\infty}\left(\begin{array}{c}
k_{4}\cdot k_{2}\\
a
\end{array}\right)(-)^{a}z_{2}^{a},\nn
(z_{3}-z_{2})^{k_{3}\cdot k_{2}} &= &
\sum_{b=0}^{\infty}\left(\begin{array}{c}
k_{3}\cdot k_{2}\\
b
\end{array}\right)(-)^{b}z_{3}^{k_{s}\cdot k_{2}-b}z_{2}^{b},\nn
(1-z_{3})^{k_{4}\cdot k_{3}} & = &
\sum_{c=0}^{\infty}\left(\begin{array}{c}
k_{4}\cdot k_{3}\\
c
\end{array}\right)(-)^{c}z_{3}^{c},~~~\label{5point-exp}
\eea
Grouping  $z_{2}$ and $z_{3}$ dependence in equation
(\ref{eq:nielson}) together we arrive
\bea A(12345)& = & \sum_{a,b,c=0}^{\infty}\,\left(\begin{array}{c}
k_{4}\cdot k_{2}\\
a
\end{array}\right)\left(\begin{array}{c}
k_{3}\cdot k_{2}\\
b
\end{array}\right)\left(\begin{array}{c}
k_{4}\cdot k_{3}\\
c
\end{array}\right)(-)^{a+b+c} \nn
& \times &
\int_{0}^{1}dz_{3}\int_{0}^{z_{3}}dz_{2}z_{3}^{k_{3}\cdot(k_{1}+k_{2})-b+c}z_{2}^{k_{1}\cdot
k_{2}+a+b}~~~\label{eq:pole-producing-int} \eea
Carrying out the integration in order, i.e., $\int dz_2$ first and
then $\int dz_3$ we obtain
\bea A(12345)& = & \sum_{a,b,c=0}^{\infty}\,\left(\begin{array}{c}
k_{4}\cdot k_{2}\\
a
\end{array}\right)\left(\begin{array}{c}
k_{3}\cdot k_{2}\\
b
\end{array}\right)\left(\begin{array}{c}
k_{4}\cdot k_{3}\\
c
\end{array}\right)(-)^{a+b+c}\nn
& \times &
\frac{2}{s_{12}+2(a+b-1)}\,\frac{2}{s_{123}+2(a+c-1)},~~~~\label{5point-exp-exp}
\eea
where  we have used $s_{12}=(k_{1}+k_{2})^{2}$,
$s_{123}=(k_{1}+k_{2}+k_{3})^{2}$, and the mass-shell conditions for
tachyons, $k_{1}^{2}=k_{2}^{2}=k_{3}^{2}=2$.

Now we consider the pole structure under the deformation
\eref{1n-deform} with pair $(1,5)$. For $s_{12}$, the poles are  located
at
\begin{equation}
z_{N}=\frac{(k_{1}+k_{2})^{2}+2(N-1)}{-q\cdot(k_{1}+k_{2})},~~~~N=a+b=0,1,...~~~~\label{5point-s12-pole}
\end{equation}
while for $s_{123}$ the poles are located at
\begin{equation}
w_{M}=\frac{(k_{1}+k_{2}+k_{3})^{2}+2(M-1)}{-q\cdot(k_{1}+k_{2}+k_{3})},~~~M=a+c=0,1,2,...
~~~~\label{5point-s123-pole}
\end{equation}
Using the BCFW recursion relation, we have
\bea A(1,2,3,4,5)=\sum_{z_{N}} {2\over s_{12}+2(N-1)} {\cal R}_{N}+
\sum_{w_{M}} {2\over s_{123}+2(M-1)} {\cal S}_{M}\eea
where ${\cal R}_{N}$ and ${\cal S}_{M}$ are corresponding residues
of poles.

~\\

{\bf Residue ${\cal R}_{N}$:} From \eref{5point-exp-exp} we can read
out the residue ${\cal R}_{N}$ as
\bea {\cal R}_{N} & = & \sum_{{\begin{array}{l}a,b=0 \\
a+b=N
\end{array}}}^\infty\sum_{c=0}^{\infty}\,\left(\begin{array}{c}
k_{4}\cdot k_{2}\\
a
\end{array}\right)\left(\begin{array}{c}
k_{3}\cdot k_{2}\\
b
\end{array}\right)\left(\begin{array}{c}
k_{4}\cdot k_{3}\\
c
\end{array}\right)(-)^{a+b+c}\left[ {2\over \WH s_{123}(z_N)+2(a+c-1)}\right]
~~~\label{RN-form-1}\eea
Noticing that
\bean \WH s_{12}(z_N)+k_3^2+2k_3\cdot \WH
k_{12}(z_N)+2(a+c-1)=2k_3\cdot \WH k_{12}(z_N)+2(c-b+1) \eean
we can rewrite
\bea \left(\begin{array}{c}
k_{4}\cdot k_{3}\\
c
\end{array}\right)(-)^{c}\left[ {2\over \WH s_{123}(z_N)+2(a+c-1)}\right] & = &
\left(\begin{array}{c}
k_{4}\cdot k_{3}\\
c
\end{array}\right)(-)^{c}\left[ {1\over k_3\cdot \WH k_{12}(z_N)+(c-b+1)}\right]\nn
& = & \sum_{c=0}^{\infty}\int_{0}^{1}dz_{3}\,
z_{3}^{k_{3}\cdot(\hat{k}_{1}+k_{2})-b+c}\left(\begin{array}{c}
k_{4}\cdot k_{3}\\
c
\end{array}\right)(-)^{c}\nn
& = & \int_{0}^{1}dz_{3}\,
z^{k_{3}\cdot(\hat{k}_{1}+k_{2})-b}\,(1-z_{3})^{k_{4}\cdot
k_{3}},~~~\label{eq:residue-123}\eea
The reason we write the sum over $c$ as the integration is clear:
the subamplitude at the right handed side should be $A(\WH P,
3,4,\WH 5)$. With this rewriting we have
\bea {\cal R}_{N} & = & \sum_{{\begin{array}{l}a,b=0 \\
a+b=N
\end{array}}}^\infty\,\left(\begin{array}{c}
k_{4}\cdot k_{2}\\
a
\end{array}\right)\left(\begin{array}{c}
k_{3}\cdot k_{2}\\
b
\end{array}\right) (-)^{N}\int_{0}^{1}dz_{3}\,
z^{k_{3}\cdot(\hat{k}_{1}+k_{2})-b}\,(1-z_{3})^{k_{4}\cdot
k_{3}}~~~\label{RN-form-2}\eea

~\\

{\bf Residue ${\cal S}_M$:} From \eref{5point-exp-exp} we can read
out the residue ${\cal S}_{M}$ as
\bea {\cal S}_{M} & = & \sum_{{\begin{array}{l}a,c=0 \\
a+c=M
\end{array}}}^\infty\sum_{b=0}^{\infty}\,\left(\begin{array}{c}
k_{4}\cdot k_{2}\\
a
\end{array}\right)\left(\begin{array}{c}
k_{3}\cdot k_{2}\\
b
\end{array}\right)\left(\begin{array}{c}
k_{4}\cdot k_{3}\\
c
\end{array}\right)(-)^{a+b+c}\left[ {2\over \WH s_{12}(w_N)+2(a+b-1)}\right]
~~~\label{SM-form-1}\eea
Using
\bea \left.\sum_{b=0}^{\infty}\frac{\left(\begin{array}{c}
k_{3}\cdot k_{2}\\
b
\end{array}\right)(-)^{b}}{\hat{k}_{1}\cdot k_{2}+\left(a+b\right)+1}\right|_{z=w_M}
& = & \sum_{b=0}^{\infty}\int_{0}^{1}dz_{2}\,
z_{2}^{\hat{k}_{1}\cdot k_{2}+a+b}\left(\begin{array}{c}
k_{3}\cdot k_{2}\\
b
\end{array}\right)(-)^{b}\nn
& = & \int_{0}^{1}dz_{2}\, z_{2}^{\hat{k}_{1}\cdot
k_{2}+a}(1-z_{2})^{k_{3}\cdot k_{2}},~~~\label{eq:residue-12}\eea
which remind us the subamplitude $A(\WH 1,2,3,\WH P)$, we get
another form
\bea {\cal S}_{M} & = & \sum_{{\begin{array}{l}a,c=0 \\
a+c=M
\end{array}}}^\infty\left(\begin{array}{c}
k_{4}\cdot k_{2}\\
a
\end{array}\right)\left(\begin{array}{c}
k_{4}\cdot k_{3}\\
c
\end{array}\right)(-)^{M}\int_{0}^{1}dz_{2}\, z_{2}^{\hat{k}_{1}\cdot
k_{2}+a}(1-z_{2})^{k_{3}\cdot k_{2}}~~~\label{SM-form-2}\eea
%

%%%%%%%%%%%%%%%%%%%%%%%%
\subsection{Four point scattering amplitude}
%%%%%%%%%%%%%%%%%%%%%%

Now we try to reproduce the same residue from the BCFW recursion
relation. To do this, we need to calculate the three point and four
point amplitudes with one general Fock state. The three point case
has been given in section \ref{example1}. Now we give the four point result.

First let us consider a simple example
\begin{equation}
\left\langle 0,k_{4}|V_{0}(k_{3},z_{3})\,
V_{0}(k_{2},z_{2})\alpha_{-m}^{\mu}|0,k_{1}\right\rangle
\end{equation}
where $V_{0}(k,z)$ stands for tachyon vertex operator (B.1) inserted
at $z$, and the initial state
$\left.\alpha_{-m}^{\mu}|0,k_{1}\right\rangle $ is raised from the
ground state by a $-m$ mode operator. Following the standard
treatment moving this mode operator to the left until it finally
annihilate the final state we obtain
\begin{equation}
(-k_{2}^{\mu}z_{2}^{m}-k_{3}^{\mu}z_{3}^{m})\left\langle
0,k_{4}|V_{0}(k_{3})\, V_{0}(k_{2})|0,k_{1}\right\rangle .
\end{equation}
In addition to all-tachyon amplitude we receive factors
$(-k_{2}^{\mu}z_{2}^{m}-k_{3}^{\mu}z_{3}^{m})$ picked up from the
commutator
\bea [:e^{ik\cdot
X(z)}:\,,\alpha_{-m}^{\mu}]=-k^{\mu}z^{m}\,\left(:e^{ik\cdot
X(z)}:\right)~. ~~~~\label{Vertex-am}\eea

For a generic normalized Fock state \eref{eq:fockstate} we repeat
the same manipulation, moving mode operators $\alpha_{-m}^{\mu}$ one
by one to the left, picking up a factor $(-k^{\mu}z^{m})$ when
passing a tachyon vertex $V(k,z)$. Putting all together we finally
have
\bea & & \left\langle 0,p|V_{0}(k_{3})\, V_{0}(k_{2})|\left\{
N_{\mu,m}\right\} ,k_{1}\right\rangle =\left\langle
0,p|V_{0}(k_{3})\,
V_{0}(k_{2})\prod_{\mu=0}^{D-1}\prod_{m=1}^{\infty}
\frac{(\alpha_{-m}^{\mu})^{N_{\mu,m}}}{\sqrt{N_{\mu,m}!m^{N_{\mu,m}}}}|0,k_{1}\right\rangle
\nn
& = &
\prod_{\mu=0}^{D-1}\prod_{m=1}^{\infty}\frac{(-k_{2}^{\mu}z_{2}^{m}-k_{3}^{\mu}z_{3}^{m})
^{N_{\mu,m}}}{\sqrt{N_{\mu,m}!m^{N_{\mu,m}}}}\left\langle
0,p|V_{0}(k_{3})\, V_{0}(k_{2})|0,k_{1}\right\rangle~~~~
\label{eq:4pt1234} \eea
where $\Spaa{0,p|V_{0}(k_{3})\, V_{0}(k_{2})|0,k_{1}}$ is known.

~\\

Similarly, if the Fock state defines the final state instead of the
initial state of an amplitude we move mode operator
$\alpha_{m}^{\mu}$ to the right hand side, yielding
\bea & &  \left\langle \left\{ N_{\mu,m}\right\}
,k_{5}|V_{0}(k_{4})\, V_{0}(k_{3})|0,p\right\rangle =\left\langle
0,k_{5}|\prod_{\mu=0}^{D-1}\prod_{m=1}^{\infty}\frac{(\alpha_{m}^{\mu})^{N_{\mu,m}}}
{\sqrt{N_{\mu,m}!m^{N_{\mu,m}}}}V_{0}(k_{4})\,
V_{0}(k_{3})|0,p\right\rangle . \nn & = &
\prod_{\mu=0}^{D-1}\prod_{m=1}^{\infty}\frac{(k_{4}^{\mu}z_{4}^{m}+k_{3}^{\mu}z_{3}^{m})
^{N_{\mu,m}}}{\sqrt{N_{\mu,m}!m^{N_{\mu,m}}}}\left\langle
0,k_{5}|V_{0}(k_{4})\, V_{0}(k_{3})|0,p\right\rangle
.~~~~\label{eq:4pt2345} \eea
It is worth to notice that the factors picked up by modes have
different signs from (\ref{eq:4pt1234}) due to the fact that
opposite signs were assigned to positive and negative modes in a
tachyon vertex operator,
\bea W_{0}=e^{\sum_{n=1}^{\infty}\frac{z^{n}}{n}\,
k\cdot\alpha_{-n}}e^{-\sum_{n=1}^{\infty}\frac{z^{n}}{n}\,
k\cdot\alpha_{n}}\eea
so that
\bea[\alpha_{m}^{\mu},:e^{ik\cdot
X(z)}:]=k^{\mu}z^{m}\,\left(:e^{ik\cdot X(z)}:\right)\eea
%

%%%%%%%%%%%%%%%%%%%%%%%%%%
\subsection{Calculation of residue ${\cal S}_M$}
%%%%%%%%%%%%%%%%%%%%%%%%%%%%

Having above preparation, we can calculate residue by summing over
immediate Fock states at given mass level $M$. In other words, at
level $M$, we should have
\begin{equation}
{\cal S}_M =\int dz_{2}\sum_{\{N_{\mu,m}\}}\left.\left\langle
0,\hat{k}_{5}|V_{0}(k_{4})|\{N_{\mu,m}\},\hat{p}\right\rangle
\left\langle \{N_{\mu,m}\},\hat{p}|V_{0}(k_{3})\,
V_{0}(k_{2})|0,\hat{k}_{1}\right\rangle
\right|_{z_{4}=z_{3}=1},~~~\label{eq:mass-level-discussion}
\end{equation}
where the summation is over modes$\{N_{\mu,m}\}$ at fixed mass level
$N=\sum_{\mu,m}\,\left(m\times\, N_{\mu,m}\right)$, so $\WH p, \WH
k_5, \WH k_1$ are all fixed by $M$.
Before giving the general discussion, let us see a few examples:
\begin{itemize}

\item {\bf Level $N=0$}: At $N=0$, $N_{\mu,m}$ must be all zero, so that equation
(\ref{eq:mass-level-discussion}) simply yields
\begin{equation}
 {\cal S}_0=\int dz_{2}\left.\left\langle
0,\hat{k}_{5}|V_{0}(k_{4})|0,\hat{p}\right\rangle \left\langle
0,\hat{p}|V_{0}(k_{3})\, V_{0}(k_{2})|0,\hat{k}_{1}\right\rangle
\right|_{z_{4}=z_{3}=1} =1\times\int_{0}^{1}dz_{2}\,
z^{k_{2}\cdot\hat{k}_{1}}(1-z_{2})^{k_{3}\cdot k_{2}},
\end{equation}
and we have an agreement with (\ref{SM-form-2}) at $a=c=0$.

\item {\bf Level $N=1$}: The $N=1$ state can only arise from states having a
single $N_{\mu,m}=1$ for $\mu=0,\dots,D-1$, while powers of other
modes remain zero
\bea {\cal S}_1 & = & \sum_{\mu,\nu}\int dz_{2}\left.\left\langle
0,\hat{k}_{5}|V_{0}(k_{4})|N_{\mu,1},\hat{p}\right\rangle
g^{\mu\nu}\left\langle N_{\nu,1},\hat{p}|V_{0}(k_{3})\,
V_{0}(k_{2})|0,\hat{k}_{1}\right\rangle \right|_{z_{4}=z_{3}=1}\nn
& = &
\int_{0}^{1}\left.dz_{2}(-k_{4})\cdot(k_{3}z_{3}+k_{2}z_{2})\,\,
z^{k_{2}\cdot\hat{k}_{1}}(1-z_{2})^{k_{3}\cdot
k_{2}}\right|_{z_{3}=1} \eea
 In addition to the usual tachyonic
Koba-Nielson formula we obtain a factor $-\left(k_{4}\cdot
k_{3}\right)z_{3}-\left(k_{4}\cdot k_{2}\right)z_{2}|_{z_{3}=1}$.
These two terms correspond to $(a,c)=(0,1)$ and $(1,0)$
respectively.

\item {\bf Level $N=2$}: The first non-trivial case happens at $N=2$. As in
the previous mass level we receive an additional term to the
tachyonic formula. For $N_{\mu,2}$ states this factor is
$\frac{-1}{2}k_{4}\cdot\left(k_{3}z_{3}^{2}+k_{2}z_{2}^{2}\right)$,
while for states with $N_{\mu_{1},1}=N_{\mu_{2},1}=1$ and
$0\leq\mu_{1}<\mu_{2}\leq D-1$ the factor is
$\frac{1}{2}\left[k_{4}\cdot\left(k_{3}z_{3}+k_{2}z_{2}\right)\right]^{2}-\frac{1}{2}
\sum_{\mu}\left[k_{4}^{\mu}(k_{3}z_{3}+k_{2}z_{2})^{\mu}\right]^{2}$,
and for states with $N_{\mu,1}=2$ we obtain
$\sum_{\mu}\left[k_{4}^{\mu}(k_{3}z_{3}+k_{2}z_{2})\right]^{2}$.
Adding all these contribution gives
\bea & &
\frac{-1}{2}k_{4}\cdot\left(k_{3}z_{3}^{2}+k_{2}z_{2}^{2}\right)+\frac{1}{2}
\left[k_{4}\cdot\left(k_{3}z_{3}+k_{2}z_{2}\right)\right]^{2}~~~\label{eq:n2-additional-terms}
\\
& = & \frac{(k_{4}\cdot k_{3})(k_{4}\cdot
k_{3}-1)}{2}z_{3}^{2}+\frac{(k_{4}\cdot k_{2})(k_{4}\cdot
k_{2}-1)}{2}z_{2}^{2}+(k_{4}\cdot k_{3})(k_{4}\cdot k_{2})z_{3}z_{2}
\nonumber  \eea
 Explicit expansion into series shows again agreement with
$\left(\begin{array}{c}
k_{4}\cdot k_{2}\\
a
\end{array}\right)\left(\begin{array}{c}
k_{4}\cdot k_{3}\\
c
\end{array}\right)(-)^{a+c}z_{2}^{a}$, with the first, second, third terms corresponding to $(a,c)=(0,2)$,
$(2,0)$ and $(1,1)$ respectively.
\end{itemize}
%

%~\\

For general level $N=\sum_{n=1}^{\infty}n\, N_{n}$ in addition to
the all-tachyon formula we have %
\footnote{Note that at every step these factors are produced in the
same pattern observed in the 4-point case, as was discussed in
appendix B, except
with $k_{3}$ now replaced by $k_{3}z_{3}^{n}+k_{2}z_{2}^{n}$.%
}
\bea & &  \sum_{\begin{array}{c}
\text{partitions of }N\\
\text{into }\{N_{n}\}
\end{array}}\,\prod_{n=1}^{\infty}\frac{\left[-k_{4}\cdot\left(
k_{3}z_{3}^{n}+k_{2}z_{2}^{n}\right)\right]^{N_{n}}}{N_{n}!\,
n^{N_{n}}}\nn
& =& \sum_{\begin{array}{c}
\text{partitions of }N\\
\text{into }\{N_{n}\}
\end{array}}\,\prod_{n}\,\sum_{N_{n}^{(2)}=0}^{\infty}\frac{\left(\begin{array}{c}
N_{n}\\
N_{n}^{(2)}
\end{array}\right)}{N_{n}!\, n^{N_{n}}}(k_{4}\cdot k_{3})^{N_{n}-N_{n}^{(2)}}z_{3}
^{n\,(N_{n}-N_{n}^{(2)})}(k_{4}\cdot k_{3})^{N_{n}^{(2)}}z_{2}^{n\,
N_{n}^{(2)}}.~~~\label{eq:intermediate-step} \eea
where in the second line above we expanded the numerator with
respect to power of $z_{2}$, which we denote as $N_{n}^{(2)}$.
Introducing the notation $N_{n}^{(3)}=N_{n}-N_{n}^{(2)}$, the
combinatorial factor can be written as
\bean \left(\begin{array}{c}
N_{n}\\
N_{n}^{(2)}
\end{array}\right)\frac{1}{N_{n}!\, n^{N_{n}}}=\frac{1}{N_{n}^{(2)}!\,
(N_{n}-N_{n}^{(2)})!\, n^{N_{n}}}=\frac{1}{N_{n}^{(2)}!\, N^{(3)}!\,
n^{N_{n}^{(2)}}n^{N_{n}^{(3)}}} \eean
Now we notice that  in equation (\ref{eq:intermediate-step}),
summing over partitions of fixed $N_{n}$ into $N_{n}^{(2)}$ and
$N_{n}^{(3)}$ first and then summing over partitions of $N$ into
$\{N_{n}\}$ secondly can be replaced by summing over partitions of
$N$ directly into $\{N_{n}^{(2)}\}$ and $\{N_{n}^{(3)}\}$, so
\eref{eq:intermediate-step} can be written as
\begin{equation}
\sum_{\text{partitions into
}N_{n}^{(2)},N_{n}^{(3)}}\,\prod_{n}\frac{1}{N_{n}^{(2)}!\,
N^{(3)}!\, n^{N_{n}^{(2)}}n^{N_{n}^{(3)}}}\left(k_{4}\cdot
k_{2}\right)^{N_{n}^{(3)}}\left(k_{4}\cdot
k_{3}\right)^{N_{n}^{(2)}}\, z_{2}^{n\, N_{n}^{(2)}}z_{3}^{n\,
N_{n}^{(3)}}.~~~\label{eq:partition-n2-n3}
\end{equation}
Defining
\bea K=\sum_{n}N_{n}^{(2)},~~ J\equiv\sum_{n}N_{n}^{(3)},~~~
a=\sum_{n}n\, N_{n}^{(2)},~~~ c=\sum_{n}n\, N_{n}^{(3)}, \eea
sum in equation (\ref{eq:partition-n2-n3}) can be divided into
summations over partitions of $\{N_{n}^{(2)}\}$ and
$\{N_{n}^{(3)}\}$ with fixed $J$, $K$, $a$, $c$ at first, and then summing
 over  $J$, $K$, and $a$\footnote{However note that $c$ should not be summed over here
because the mass level $(a+c)=\sum_{n}n(N_{n}^{(2)}+N_{n}^{(3)})=N$
is understood as a fixed number at every pole. }, i.e., equation
(\ref{eq:partition-n2-n3}) is equal to
\begin{equation}
\sum_{a}\sum_{J,K}\frac{S(c,J)}{c!}\,\frac{S(a,K)}{a!}\left(k_{4}\cdot
k_{2}\right)^{J}\left(k_{4}\cdot k_{3}\right)^{K}\,
z_{2}^{a}z_{3}^{c},~~~\label{eq:striling-express-5pt}
\end{equation}
where  Striling numbers of the first kind are given by
\begin{equation}
S(a,K)=\sum_{\text{partitions
}N_{n}^{(2)}}\,\frac{a!}{N_{n}^{(2)}!\, n^{N_{n}^{(2)}}},~~~~
S(c,J)=\sum_{\text{partitions
}N_{n}^{(3)}}\,\frac{c!}{N_{n}^{(3)}!\, n^{N_{n}^{(3)}}},
\end{equation}

Now we are almost done. Summing equation
(\ref{eq:striling-express-5pt}) over $J$ and $K$ yields
\begin{equation}
\left(\begin{array}{c}
k_{4}\cdot k_{2}\\
a
\end{array}\right)\left(\begin{array}{c}
k_{4}\cdot k_{3}\\
c
\end{array}\right)(-)^{a+c}\, z_{2}^{a}\, z_{3}^{c}.
\end{equation}
Inserting the result back into (\ref{eq:mass-level-discussion}) we
see that
\bea {\cal S}_M & &  \int dz_{2}\left.\left\langle
0,\hat{k}_{5}|V_{0}(k_{4})|\{N_{\mu,m}\},\hat{p}\right\rangle
\left\langle \{N_{\mu,m}\},\hat{p}|V_{0}(k_{3})\,
V_{0}(k_{2})|0,\hat{k}_{1}\right\rangle \right|_{z_{4}=z_{3}=1} \nn
& = & \sum_{a}\int dz_{2}\left\langle
0,\hat{k}_{5}|V_{0}(k_{4})|0,\hat{p}\right\rangle \left\langle
0,\hat{p}|V_{0}(k_{3})\, V_{0}(k_{2})|0,\hat{k}_{1}\right\rangle \nn
& &  \times\left.\left(\begin{array}{c}
k_{4}\cdot k_{2}\\
a
\end{array}\right)\left(\begin{array}{c}
k_{4}\cdot k_{3}\\
c
\end{array}\right)(-)^{a+c}\, z_{2}^{a}\, z_{3}^{c}\right|_{z_{4}=z_{3}=1}
\nn
& = & \sum_{a=0, a+c=M}^{M}\left(\begin{array}{c}
k_{4}\cdot k_{2}\\
a
\end{array}\right)\left(\begin{array}{c}
k_{4}\cdot k_{3}\\
c
\end{array}\right)(-)^{a+c}\int_{0}^{1}dz_{2}\, z_{2}^{\hat{k}_{1}\cdot k_{2}+a}(1-z_{2})^{k_{3}\cdot k_{2}},
\eea
which is the form \eref{SM-form-2} we want to prove.

The other residue ${\cal R}_N$  can be derived from BCFW
prescription following similar procedures.

%%%%%%%%%%%%%%%%%%%%%%%%%%
\section{The general proof}
%%%%%%%%%%%%%%%%%%%%%%%%%%%
\label{sec-n-pt}
Having done above two examples, we would like to have a general
understanding. The method we will use in this section will be a
little different although it is easy to translate languages between
these two approaches.

%%%%%%%%%%%%%%%%%%%%%%%%
\subsection{String theory calculation}
%%%%%%%%%%%%%%%%%%%%%%%%

In open string theory, the ordered tree-level amplitude is given by
\bea A_M & = & g^{M-2} \int \delta(y_A- y_A^0) \delta(
y_B-y_B^0)\delta(y_c-y_c^0) (y_A-y_B) (y_A- y_C) (y_B-y_C) \nn & &
\prod_{i=2}^M \theta(y_{i-1}-y_i) \prod_{j=1}^M d y_j
\Spaa{0;0\left| {V(k_1,y_1)\over y_1}...{V(k_M,y_M)\over
y_M}\right|0;0}~~~~\label{M-point-form-1}\eea
Using three delta-function, we can take $y_M=0, y_2=1, y_1=\infty$,
so the amplitude can be written as
\bea A_M & = & g^{M-2} \int_0^1 d y_3 \int_0^{y_3} d
y_4....\int_0^{y_{M-2}} d y_{M-1} \Spaa{\phi_1(k_1)\left|
{V(k_2,1)}{V(k_3,y_3)\over y_3}...{V(k_{M-1},y_{M-1})\over
y_{M-1}}\right|\phi_M(k_M)}~~~~\label{M-point-form-2}\eea
where we have used the definition of initial state and final state
\bea \ket{\Lambda;k}=\lim_{y\to 0} {V_{\Lambda}(k,y)\over
y}\ket{0;0},~~~~~\bra{\Lambda;k}=\lim_{y\to \infty}
{yV_{\Lambda}(k,y)}\ket{0;0}\eea
Next we define $y_i=z_3 z_4...z_i$ with $i=3,...,M-1$, from which we
can solve
\bea z_3=y_3,~~~~z_i={y_i\over y_{i-1}},~~~i=4,...,M-1\eea
Now let us fix all $y_i$ except transform $y_{M-1}= z_{M-1}
y_{M-2}$, then using
\bea V_{\Lambda}(k,z)= z^{L_0}  V_{\Lambda}(k,z=1) z^{-L_0}\eea
we get
\bean ....\int_0^1 d z_{M-1} y_{M-2} y_{M-1}^{L_0-2} V(k_{M-1},1)
y_{M-1}^{-L_0+1}\ket{\phi_M(k_M)}=...\left[\int_0^1 d z_{M-1}
y_{M-2}^{L_0-1} z_{M-1}^{L_0-2} \right]V(k_{M-1},1)
\ket{\phi_M(k_M)}\eean
where we have used the physical condition $(L_0-1)\ket{\phi_M}=0$.
Now we change $y_{M-2}= z_{M-2} y_{M-3}$, then we have
\bean & & ...\int_0^1 dz_{M-2} y_{M-3}  y_{M-2}^{L_0} {
V(k_{M-2},1)\over y_{M-2}} y_{M-2}^{-L_0}\int_0^1 d z_{M-1}
y_{M-2}^{L_0-1} z_{M-1}^{L_0-2} V(k_{M-1},1) \ket{\phi_M(k_M)}
\nn
& = & ...\int_0^1 dz_{M-2} y_{M-3}  y_{M-2}^{L_0-2} { V(k_{M-2},1)}
\int_0^1 d z_{M-1}  z_{M-1}^{L_0-2} V(k_{M-1},1)
\ket{\phi_M(k_M)}\nn
& = & ...\left[\int_0^1 dz_{M-2} y_{M-3}^{L_0-1}  z_{M-2}^{L_0-2}
\right]{ V(k_{M-2},1)} {1\over L_0-1} V(k_{M-1},1)
\ket{\phi_M(k_M)}\eean
where we have used $\int_0^1 dz z^{L_0-2}={1\over L_0-1}$ is the
string propagator.

Comparing expressions from last two steps, we see that we can
iterate this procedure to
\bea A_M & = & g^{M-2} \Spaa{\phi_1 \left|V_2(k_2){1\over L_0-1}
V_3(k_3)...{1\over L_0-1}
V_{M-1}(k_{M-1})\right|\phi_M}~~~~\label{M-point-form-3}\eea
Form \eref{M-point-form-3} is the convenient one to compare with
BCFW recursion relation, because locations of poles are clearly
indicated by propagator ${1\over L_0-1}$. For example, for ${1\over
L_0-1}$ between vertex operators $V_{i}$ and $V_{i+1}$, pole
locations are given by
\bea {1\over 2} (k_1+...+k_i)^2+N-1=0,~~~~N=0,1,2,...
~~~\label{gen-pole}\eea

Now let us consider the  $(1,M)$-deformation given in
\eref{1n-deform} and use $z_{iN}$ to indicate the solution obtained
from equation \eref{gen-pole} with $k_1\to k_1+zq$. Because it has
been proved that boundary contribution is zero under the deformation
at least for some kinematic region, we have immediately
\bea A_M & = & g^{M-2} \sum_{i=2}^{M-2} \sum_{N=0}^\infty { 2 {\cal
R}_{i,N} \over (k_1+k_2+...+k_i)^2+2(N-1)}~~~~\label{String-Exp}\eea
where
\bea {\cal R}_{i,N} & = & \Spaa{\Phi_{i,N}|\Psi_{i,N}}
\nn
\bra{\Phi_{i,N}} & = & \bra{\phi_1(k_1+z_{i,N} q)|
\left|V_2(k_2){1\over L_0-1} V_3(k_3)...{1\over
L_0-1}V_i(k_i)\right.} \nn
\ket{\Psi_{i,N}}& = & \ket{\left. V_{i+1}(k_{i+1}) {1\over L_0-1}...
V_{M-1}(k_{M-1})\right|\phi_M(k_M-z_{i,N} q)}
~~~~\label{String-residue}\eea
What we want to prove is that residue ${\cal R}_{i,N}$ can be
obtained from summing over intermediate physical states prescribed by BCFW
on-shell recursion relation.

%%%%%%%%%%%%%%%%%%%%%
\subsection{The proof}
%%%%%%%%%%%%%%%%%%%%%%

Now we give our proof. First, we notice that both states
$\bra{\Phi_{i,N}}, \ket{\Psi_{i,N}}$ are physical
states\footnote{The proof  can be found in a standard text, for
example in \textit{Superstring Theory} by Green, Schwarz and Witten\cite{String}
(chapter 7, vol. 1.).}, thus in the frame work of
DDF-state construction, both physical states can be written as
$\ket{s_{phy}}+\ket{f}$, where $\ket{f}$ is the DDF-state while
$\ket{s_{phy}}$ is physical spurious states. Using the property of
spurious state, we have
\bea \Spaa{\Phi_{i,N}|\Psi_{i,N}} & = & \Spaa{ s_{i,N}^L+
f^L_{i,N}|s_{i,N}^R + f^R_{i,N}}= \Spaa{ f^L_{i,N}|
f^R_{i,N}}~~~\label{LR-inner-1}\eea

Having established \eref{LR-inner-1} we insert identity operator in
the Fock space with  given momentum
$P_{i,N}=k_1+z_{i,N}q+k_2+...+k_i$ and annihilated by $(L_0-1)$, so
\bea \Spaa{ f^L_{i,N}| f^R_{i,N}} & = & \sum_{i} \Spaa{
f^L_{i,N}|\psi_i^\dagger(P_{i,N})} \Spaa{\psi_{i}(P_{i,N})|
f^R_{i,N}}~~~\label{LR-inner-2}\eea
where set $\{\ket{\psi_{i}(P_{i,N})}\}$ can be  any  normalized
orthogonal basis. In DDF-frame work, a general state can be written
as the linear combination of $\ket{k},\ket{s},\ket{f}$, i.e., a
choice of the basis is $\ket{k},\ket{s},\ket{f}$. Using the
definition of states, we see immediately that $\Spaa{s|f}=0$ and
$\Spaa{k|f}=0$, thus
\bea \Spaa{ f^L_{i,N}| f^R_{i,N}} & = & \sum_{i} \Spaa{
f^L_{i,N}|f_i^\dagger(P_{i,N})} \Spaa{f_{i}(P_{i,N})|
f^R_{i,N}}\nn
& = & \sum_i \Spaa{ s_{i,N}^L+ f^L_{i,N}|f_i^\dagger(P_{i,N})}
\Spaa{f_{i}(P_{i,N})|s_{i,N}^R +
f^R_{i,N}}\nn
& = & \sum_i \Spaa{\Phi_{i,N}|f_i^\dagger(P_{i,N})}
\Spaa{f_{i}(P_{i,N})|\Psi_{i,N}}~~~\label{LR-inner-3}\eea
Using \eref{LR-inner-1} and \eref{LR-inner-3} we see immediately
\bea {\cal R}_{i,N}= \sum_i \Spaa{\Phi_{i,N}|f_i^\dagger(P_{i,N})}
\Spaa{f_{i}(P_{i,N})|\Psi_{i,N}}~~~\label{BCFW-residue} \eea
which is the prescription given by BCFW recursion relation. Thus we
have given our proof.

%%%%%%%%%%%%%%%%%%%%%%
\subsection{Practical method for summing over physical states}
%%%%%%%%%%%%%%%%%%%%%%

Having shown that BCFW recursion relation gives the right string
amplitude, we need to explain how to sum over physical states. The
difficulty of the sum is that the physical state is hard to describe
in general, i.e., we do not know how to write down polarization
vector for a given physical state. However, from the equivalent
between \eref{LR-inner-2} and \eref{LR-inner-3} we see that we can
replace the sum over all physical states to the sum over whole Fock
space with given momentum and annihilated by $(L_0-1)$. For the Fock
space, there is a freedom with the choice of basis and the one
convenient for real calculation is oscillation basis defined in
\eref{eq:fockstate}. Thus the residue can be calculated by
\bea {\cal R}_{i,N}&= & \sum_{\{N_{\mu,n}\} }
\Spaa{\Phi_{i,N}|\{N_{\mu,n}\};\WH P} {\cal T}_{\{N_{\mu,n}\}}
\Spaa{\{N_{\mu,n}\};\WH P|\Psi_{i,N}}\nn
& = & \sum_{\sum_n nN_{n}=N }
\Spaa{\Phi_{i,N}\left|\left\{\prod_{n=1}^\infty {
(\a_{-n}^{{\mu_{N_{n},1}}} \a_{-n}^{{\mu_{N_{n},2}}}...
\a_{-n}^{{\mu_{N_{n},N_{n}}}})\over \sqrt{N_{n}!
n^{N_{n}}}}\right\}\right|0;\WH P} \prod_{n=1}^\infty
(g_{{\mu_{N_{n},1}}{\nu_{N_{n},1}}}
g_{{\mu_{N_{n},2}}{\nu_{N_{n},2}}}...
g_{{\mu_{N_{n},N_{n}}}{\nu_{N_{n},N_{n}}}})\nn
& & \Spaa{0;\WH P\left| \left\{\prod_{n=1}^\infty {
(\a_{+n}^{{\nu_{N_{n},1}}} \a_{+n}^{{\nu_{N_{n},2}}}...
\a_{+n}^{{\nu_{N_{n},N_{n}}}})\over \sqrt{N_{n}!
n^{N_{n}}}}\right\}\right|\Psi_{i.N}}~~~\label{Cal-res}\eea
%

%%%%%%%%%%%%%%%%%%%%%%
\section{Scattering with higher spin particles}
\label{sec-vector} %%%%%%%%%%%%%%%%%%

Having established the general method given in \eref{Cal-res},
let us consider scatterings when higher spin particles are
present. However,
before doing this, let us recall some results coming from scattering
amplitudes of pure tachyons. 
By checking with \eref{4-pt-res} and \eref{eq:striling-express-5pt},
we see that residues are given as series of Lorentz invariants
$k_{i}\cdot k_{j}$ with  coefficients  given by Stirling number of
the first kind
$s(N,J)=\sum_{\{N_{n}\}}\prod_{n=1}^{\infty}\frac{1}{N_{n}!n^{N_{n}}}$.
Summing over powers of $k_{i}\cdot k_{j}$ reproduces the residue in
combinatorial form observed in \cite{Cheung:2010vn}. This relation
is established by writing generating function of Stirling number
into two different forms
\bea
  e^{X\, ln(1-z)} & = & e^{-X\,(z+\frac{z^{2}}{2}+\frac{z^{3}}{3}+\dots)}
=e^{-X\, z}e^{-X\,\frac{z^{2}}{2}}e^{-X\frac{z^{3}}{3}}\dots \nn & =
&
\left(1+(-)Xz+\frac{(-)^{2}}{2!}X^{2}z^{2}+\dots\right)\,\left(1+(-)X\frac{z^{2}}{2}+(-)^{2}X\left(\frac{z^{2}}{2}\right)^{2}+\dots\right)\dots
  ~~~~\label{eq:expansion-gen-funct}
\eea
and
\bea (1-z)^{X} & =
&\sum_{a=1}^{\infty}(-)^{a}\frac{s(a,J)}{a!}X^{J}z^{a} =
\sum_{a}(-)^{a}\left(\begin{array}{c}
X\\
a
\end{array}\right)z^{a}
\eea
by matching power of $z$ and setting $X=k_2\cdot k_3$. In fact, it
is straightforward to see that residues in an arbitrary $n$-point
pure tachyon scattering amplitude can be read off from products of
generating functions
\begin{equation}
e^{X_{23}\, ln(1-z_{23})}e^{X_{24}\, ln(1-z_{24})}\dots
e^{X_{n-2,n-1}\, ln(1-z_{n-2,n-1})}
\end{equation}
with $X_{ij}=k_{i}\cdot k_{j}$, $z_{ij}=z_{j}/z_{i}$, and residues
in tachyonic recursion relation can be found through binomial
expansion of
\begin{equation}
(1-z_{23})^{X_{23}}\,(1-z_{24})^{X_{24}}\dots(1-z_{n-2,n-1})^{X_{n-2,n-1}}.
\end{equation}

Having recalled the experience from tachyon amplitude, now we discuss
the scattering amplitude of 3-tachyon and 1-vector, which  is given by
\bea
A(1 \O2 34) & = &  \left.\int_{0}^{1}\frac{dz_{2}}{z_{2}}\,\left\langle 0,k_{1}\left|\,\left( \epsilon_{2} \cdot \dot{X} \,:e^{ik_{2}\cdot X(z_{2})}:\right)\,\left(:e^{ik_{3} \cdot X(z_{3})}:\right) \,\right|0,k_{4}\right\rangle \right|_{z_{3}=1}  ~~~~\label{eq:a-1234}
\\
&= &\int_{0}^{1} \, dz_{2} \, \left( - \epsilon_{2} \cdot k_{1}(1-z_{2})^{k_{3} \cdot k_{2}}\,z_{2}^{k_{1} \cdot k_{2}-1}+ \epsilon_{2} \cdot k_{3} \, (1-z_{2})^{k_{2} \cdot k_{3}-1} \, z_{2}^{k_{1} \cdot k_{2}} \right) 
\eea
where $\O 2$ means that the second particle is a vector.  As in the case of
pure tachyon scattering we binomially expanding
$(1-z_{2})^{k_{3}\cdot k_{2}}$ in (\ref{eq:a-1234}) and integrating
over $z_{2}$, yielding
\bea  A(1 \O2 34) & & = -\sum_{a=0}^{\infty} (-)^{a} \epsilon_{2}\cdot k_{1}\left(\begin{array}{c}
k_{3}\cdot k_{2}\\
a
\end{array}\right)\frac{2}{(k_{2}+ k_{1})^{2}+2(a-1)} \nn
& &
+ \sum_{a=1}^{\infty} (-)^{a-1} \epsilon_{2}\cdot k_{3}\left(\begin{array}{c}
k_{3}\cdot k_{2}-1\\
a -1
\end{array}\right)\frac{2}{(k_{2}+ k_{1})^{2}+2(a-1)}.
 ~~~\label{eq:scalar-like-1v3t}
\eea
We are  interested in relating residue in
(\ref{eq:scalar-like-1v3t}) with residue given by BCFW prescription
\begin{equation}
\left.\left\langle 0,k_{1}\left|\,\epsilon_{2}\cdot\dot{X}\,:e^{ik_{2}\cdot X(z_{2})}:\left|\{N_{\mu,m}\},p\right\rangle \mathcal{T}_{\{N_{\mu,m}\}}\left\langle \{N_{\mu,m}\},p\right|:e^{ik_{3}\cdot X(z_{3})}:\,\right|0,k_{4}\right\rangle \right|_{z_{2}=z_{3}=1}.\label{eq:3-pts}
\end{equation}
It is straightforward to see at the first few levels, residues in
(\ref{eq:scalar-like-1v3t}) agree with those prescribed by
(\ref{eq:3-pts})  table \ref{tab:2}.
\begin{table}[h]
\centering
\begin{tabular}{|c|c|c|}
\hline
 & intermediate state $\left|\{N_{\mu,m}\}\right\rangle \mathcal{T}_{\{N_{\mu,m}\}}\left\langle \{N_{\mu,m}\}\right|$ & contribution $\sim\epsilon_{2} \cdot k_{3}$\tabularnewline
\hline \hline $N=0$ & $\left|0\right\rangle \left\langle 0\right|$ &
absent\tabularnewline \hline $N=1$ &
$\frac{\alpha_{-1}^{\mu}}{\sqrt{1}}\left|0\right\rangle
\eta_{\mu\nu}\left\langle 0\right|\frac{\alpha_{1}^{\nu}}{\sqrt{1}}$
& $(-)\left(\epsilon_{2} \cdot k_{3}\right)$\tabularnewline \hline
$N=2$ & $\begin{array}{c}
\frac{\alpha_{-2}^{\mu}}{\sqrt{2}}\left|0\right\rangle \eta_{\mu\nu}\left\langle 0\right|\frac{\alpha_{2}^{\nu}}{\sqrt{2}}\\
\sum_{\mu_{1}<\mu_{2}}\,\frac{\alpha_{-1}^{\mu_{1}}}{\sqrt{1}}\,\frac{\alpha_{-1}^{\mu_{2}}}{\sqrt{1}}\left|0\right\rangle \eta_{\mu_{1}\nu_{1}}\eta_{\mu_{2}\nu_{2}}\left\langle 0\right|\frac{\alpha_{1}^{\nu}}{\sqrt{1}}\frac{\alpha_{1}^{\nu_{2}}}{\sqrt{1}}\\
\frac{1}{\sqrt{2!}}\,\frac{\alpha_{-1}^{\mu}}{\sqrt{1}}\,\frac{\alpha_{-1}^{\mu}}{\sqrt{1}}\left|0\right\rangle
\left(\eta_{\mu\nu}\right)^{2}\left\langle
0\right|\frac{1}{\sqrt{2!}}\,\frac{\alpha_{1}^{\nu}}{\sqrt{1}}\frac{\alpha_{1}^{\nu}}{\sqrt{1}}
\end{array}$ & $\begin{array}{c}
(-)\left(\epsilon_{2} \cdot k_{3}\right)\\
\left(\epsilon_{2} \cdot k_{3}\right)\left(k_{3}\cdot k_{2}\right)\\
\\
\end{array}$\tabularnewline
\hline
\end{tabular}
\caption{Residues of $3$-tachyon, $1$-vector scattering for first
three levels \label{tab:2}}
\end{table}

 Note that algebraically, the
first term proportional to $\epsilon_{2} \cdot k_{1}$ in
(\ref{eq:scalar-like-1v3t}) was obtained from moving an operator
$\epsilon_{2} \cdot\alpha_{0}$ in
$\epsilon_{2} \cdot\dot{X}(z_{2})=\epsilon_{2} \cdot(\alpha_{-1}z_{2}^{1}+\dots+\alpha_{0}z_{2}^{0}+\alpha_{1}z_{2}^{-1}+\dots)$
to the left, acting upon final state $\left|0,k_{1}\right\rangle $
in the standard process of normal ordering, which simply reproduces
the pure tachyon residue since rest of its kinematic dependence was
contributed from $\left\langle 0,k_{1}\left|:e^{ik_{2}\cdot
X(z_{2})}:\,:e^{ik_{3}\cdot X(z_{3})}:\right|0,k_{4}\right\rangle $.
It is therefore straightforward to show that, following the same
expansion as in the case of pure tachyon scattering, at each mass
level residue contributed from this term is connected to BCFW
prescription by generating function for Stirling number of the first
kind. New structure however, is found in the second term
proportional to $\epsilon_{2} \cdot k_{3}$ in
(\ref{eq:scalar-like-1v3t}), which was produced by moving positive
mode operators $\alpha_{1}z_{2}^{-1}+\alpha_{2}z_{2}^{-2}+\dots$ in
$\epsilon_{2} \cdot\dot{X}(z_{2})=\epsilon_{2} \cdot(\alpha_{-1}z_{2}^{1}+\dots+\alpha_{0}z_{2}^{0}+\alpha_{1}z_{2}^{-1}+\dots)$
to the right and contracting with intermediate states. For example
when we have a Fock state
$\frac{\alpha_{-q}^{\mu_{1}}\alpha_{-r}^{\mu_{2}}}{\sqrt{q}\sqrt{r}}\,\frac{1}{\sqrt{2!}}\left|0,p\right\rangle
$ as intermediate state, equation (\ref{eq:3-pts}) reads
\begin{equation}
\left.\left\langle 0,k_{1}\left|(\epsilon_{2} \cdot\sum_{n=1}^{\infty}\alpha_{n}\, z_{2}^{-n})e^{-\frac{1}{n}k_{2}\cdot\alpha_{n}z_{2}^{-n}}\frac{\alpha_{-q}^{\mu_{1}}\alpha_{-r}^{\mu_{2}}}{\sqrt{q}\sqrt{r}}\,\frac{1}{\sqrt{2!}}\right|0,p\right\rangle \eta_{\mu_{1}\mu_{2}}\eta_{\nu_{1}\nu_{2}}\left\langle 0,p\left|\frac{\alpha_{q}^{\nu_{1}}\alpha_{r}^{\nu_{2}}}{\sqrt{q}\sqrt{r}}\,\frac{1}{\sqrt{2!}}e^{-\frac{1}{n}k_{3}\cdot\alpha_{n}z_{3}^{n}}\right|0,k_{4}\right\rangle \right|_{z_{2}=z_{3}=1}.
\end{equation}
Contribution proportional to $\epsilon_{2} \cdot k_{3}$ is produced
by contracting an $\alpha_{q}$ or $\alpha_{r}$ in $\epsilon_{2} \cdot\dot{X}(z_{2})$
with Fock state, yielding
\begin{equation}
\left.\frac{(\epsilon_{2} \cdot k_{3})\times q}{q}(\frac{z_{3}}{z_{2}})^{q}\,\frac{(k_{3}\cdot k_{2})}{r}(\frac{z_{3}}{z_{2}})^{r}+\frac{(k_{3}\cdot k_{2})}{q}(\frac{z_{3}}{z_{2}})^{q}\,\frac{(\epsilon_{2} \cdot k_{3})\times r}{r}(\frac{z_{3}}{z_{2}})^{r}\right|_{z_{2}=z_{3}=1}.
\end{equation}
Therefore generically residue (\ref{eq:3-pts}) proportional to $\epsilon_{2} \cdot k_{3}$
at level $N=a$ is given by $z^{a}$ term expansion coefficient of the derivative of generating function
\bea
&& \frac{(\epsilon_{2} \cdot k_{3})}{(k_{2}\cdot k_{3})}\, z\frac{d}{dz}e^{(k_{2}\cdot k_{3})\, \mathrm{ln}(1-z)} ~~~~\label{eq:gen-func-vect}
\\
& =&\frac{(\epsilon_{2} \cdot k_{3})}{(k_{2}\cdot k_{3})}\, z\frac{d}{dz}\left(e^{-X\, z}e^{-X\,\frac{z^{2}}{2}}e^{-X\frac{z^{3}}{3}}\dots\right). \nonumber
\eea
Note that we may as well express the generating function (\ref{eq:gen-func-vect})
above as
\begin{equation}
(\epsilon_{2} \cdot k_{3}) \,   z\frac{d}{dz} \left[ \mathrm{ln}(1-z)\right]  e^{(k_{2}\cdot k_{3})\, \mathrm{ln}(1-z)} \, ,
~~~\label{vt-gen-funct2}
\end{equation}
from which it is obvious that BCFW prescription yields the same residue as tachyonic recursion relation of $1$-vector $3$-tachyon amplitude, since the tachyonic recursion relation was
derived from binomial expansion of standard worldsheet integral
formula that takes the same form as (\ref{vt-gen-funct2}). 
\\
\\
%%%%%%%%%%%%%%%%%
%%%%%%%%%%%%%%%%%
\subsection*{Explicit recursion relation}
%%%%%%%%%%%%%%%%

Here we present an explicit calculation of the term proportional to $\epsilon_{2} \cdot k_{3}$ in Eq.(6.7). By using Eq.(6.8), the term
proportional to $\epsilon_{2}\cdot k_{3}$ with mass level $N$ can be
calculated by gluing two 3-point functions\newline
\begin{equation*}
I_{N}=\sum\limits_{\left\{ \sum {mN_{m}}=N\right\} }{\Big \langle\,k_{1};0\,%
\Big |\,\Big (\sum\limits_{n=1}^{\infty }{\epsilon_{2} \cdot \alpha _{n}}\Big )%
\,V_{0}(k_{2})\,\Big |\left\{ N_{m}\right\} ;P\Big \rangle\,\mathcal{T}%
_{\left\{ N_{m}\right\} }\,\Big \langle\left\{ N_{m}\right\} ;P\,\Big |%
\,V_{0}(k_{3})\,\Big |\,k_{4};0\ \Big \rangle}\,\Bigg |_{z_{2}=1}.
\end{equation*}%
\newline
For convenience, let us denote the two 3-point functions as \newline
\begin{align}
A_{L}& =A_{L}(k_{1},k_{2},P)=\Big \langle\,k_{1};0\,\Big |\,\Big (%
\sum\limits_{n=1}^{\infty }{\epsilon_{2} \cdot \alpha _{n}}\Big )\,V_{0}(k_{2})\,%
\Big |\left\{ N_{m}\right\} ;P\Big \rangle\,\Big |_{z_{2}=1}, \\
A_{R}& =A_{R}(P,k_{3},k_{4})=\Big \langle\left\{ N_{m}\right\} ;P\,\Big |%
\,V_{0}(k_{3})\,\Big |\,k_{4};0\ \Big \rangle\,\Big |_{z_{2}=1}.
\end{align}%
\newline
The term $A_{R}$ was obtained in Eq.(3.29) previously, while $A_{L}$ can be
calculated to be (we ignore the momentum dependent part) \newline
\begin{align}
A_{L}& =\sum\limits_{n=1}^{\infty }\Big \langle\,0\,\Big |\,\Big (\epsilon_{2}\cdot \alpha _{n}\Big )\,\prod_{m=1}^{\infty }{e^{-\frac{k_{2}\cdot
\alpha _{m}}{m}}\frac{\big (\alpha _{-m}^{\mu }\big )^{N_{m}}}{\sqrt{%
m^{N_{m}}\,N_{m}!}}}~\Big |0\Big \rangle \\
& =\sum\limits_{n=1}^{\infty }\Big \langle\,0\,\Big |\,\Big (\epsilon_{2}\cdot \alpha _{n}\Big )\,\Bigg [\,e^{-\frac{k_{2}\cdot \alpha _{n}}{n}}%
\frac{\big (\alpha _{-n}^{\mu }\big )^{N_{n}}}{\sqrt{n^{N_{n}}\,N_{m}!}}%
\Bigg ]\prod_{m=1,m\neq n}^{\infty }{e^{-\frac{k_{2}\cdot \alpha _{m}}{m}}%
\frac{\big (\alpha _{-m}^{\mu }\big )^{N_{m}}}{\sqrt{m^{N_{m}}\,N_{m}!}}}~%
\Big |0\Big \rangle.
\end{align}%
\newline
In the presence of $\epsilon_{2}\cdot \alpha _{n}$ term, one notes that
only term of order $(N_{n}-1)$ in the Taylor expansion of $\mathrm{exp}\big[%
\,-k_{2}\cdot \alpha _{n}/n~\big]$ inside the square bracket will
contribute. By using $[\,\alpha _{m}^{\mu }\,,\,\alpha _{n}^{\nu
}\,]=m\delta _{m+n}\eta _{\mu \nu }$ , we get \newline
\begin{equation}
A_{L}=\sum\limits_{n=1}^{\infty }\,\left\{ \Bigg [\,\frac{%
(-)^{N_{n}-1}\,n\,N_{n}\,\epsilon_{2}^{\mu }\,(k_{2}^{\mu })^{N_{n}-1}}{%
\sqrt{\,n^{N_{n}}\,N_{n}!}}\Bigg ]\prod_{m=1,m\neq n}^{\infty }{\,\frac{\big
(-k_{2}^{\mu }\big )^{N_{m}}}{\sqrt{m^{N_{m}}\,N_{m}!}}}\right\} .
\end{equation}%
\newline
Combining $A_{R}$ and $A_{L}$ and summing over all states with $\sum_{m}{%
mN_{m}}=N$ yields \newline
\begin{equation}
I_{N}=\sum\limits_{\left\{ N=\sum_{m}{mN_{m}}\right\} }(-)\,N\,\frac{%
\,\epsilon_{2}\cdot k_{3}}{k_{2}\cdot k_{3}}\prod\limits_{m=1}^{\infty }{%
\frac{\big (-k_{2}\cdot k_{3}\big )^{N_{m}}}{m^{N_{m}}\,N_{m}!}.}
\end{equation}%
\newline
We can now use the definition of Stirling number of the first kind to get%
\begin{equation}
I_{N}=\,\epsilon_{2}\cdot k_{3}~\sum\limits_{J=1}^{N}\,\frac{s(N,J)}{N!}%
\,(-)^{N-1}\,N\,(k_{2}\cdot k_{3})^{J-1}.
\end{equation}%
\newline
Finally the expression can be further reduced to%
\begin{equation}
I_{N}=\,\epsilon _{2}\cdot k_{3}~(-)^{N-1}\,{\binom{k_{2}\cdot k_{3}-1}{N-1}.%
}
\end{equation}
\\~
%%%%%%%%%%%%%%%%%%%%
%%%%%%%%%%%%%%%%%%%%

In the following, instead of the operator method adopted previously, we will
use path-integral approach \cite{KLT} to calculate the generating function
for the rank-two tensor, three tachyons amplitude. As a warm up exercise, we
first use this method to rederive Eq.(6.12) for the vector, three tachyons
amplitude. We first note that the amplitude can be written as%
\begin{eqnarray}
\mathcal{A} &=&\int \prod_{i=1}^{1}dz_{i}<e^{ik_{1}X(z_{1})}\epsilon
_{2}\cdot \partial
X(z_{2})e^{ik_{2}X(z_{3})}e^{ik_{3}X(z_{3})}e^{ik_{4}X(z_{4})}>  \label{1} \\
&=&\int
\prod_{i=1}^{4}dz_{i}<e^{ik_{1}X(z_{1})}e^{ik_{2}X(z_{2})+i\epsilon
_{2}\cdot \partial X(z_{2})}e^{ik_{3}X(z_{3})}e^{ik_{4}X(z_{4})}>\mid _{%
\text{linear in }\epsilon _{2}}  \label{2} \\
&=&\int \prod_{i=1}^{4}dz_{i}\text{ }\exp{[-\sum_{l<j}k_{l\mu }k_{j\nu
}<X^{\mu }(z_{l})X^{\nu }(z_{j})>-\sum_{j\neq 2}\epsilon _{2\mu }k_{j\nu
}<\partial X^{\mu }(z_{l})X^{\nu }(z_{j})>]}\mid _{\text{linear in }\epsilon
_{2}}  \label{3} \\
&=&\int_{0}^{1}dz(1-z)^{k_{2}\cdot k_{3}}z^{k_{1}\cdot k_{2}}\left[\frac{%
\epsilon _{2}\cdot k_{1}}{z}-\frac{\epsilon _{2}\cdot k_{3}}{1-z}\right].
\label{4}
\end{eqnarray}%
In the last equality, we have used the worldsheet $SL(2,R)$ to set the
positions of the four vertex at $0,z,1$ and $\infty $, and the propagator $%
<X^{\mu }(z_{l})X^{\nu }(z_{j})>=-\eta ^{\mu \nu }\ln (z_{l}-z_{j})$. Note
that the term proportional to $\epsilon _{2}\cdot k_{1}$ has been
considered previously for the calculation of four tachyons amplitude. One
can now see from Eq.(\ref{3}) that the generating function for amplitude
proportional to the term $\epsilon _{2}\cdot k_{3}$ is%
\begin{eqnarray}
G_{1} &=&\exp ^{\left\{ -k_{3}\cdot k_{2}[-\ln (1-z)]\right\} }\exp ^{\left\{ -\epsilon _{2}\cdot k_{3}z\frac{d}{dz}[-\ln (1-z)]\right\} }\mid _{%
\text{linear in }\epsilon_{2}}  \label{5} \\
&=&(\epsilon _{2}\cdot k_{3})z\frac{d}{dz}[\ln (1-z)]\exp^{{\lbrace
k_{3}\cdot k_{2}[\ln (1-z)]\rbrace } } \label{6}
\end{eqnarray}%
which is the same with Eq.(6.12). Therefore the derivative of generating
function in Eq.(6.11) can be traced back to the derivative part $\partial
X^{\mu }$ of the vector vertex.  We now generalize the calculation to the
higher spin cases. For example, for the spin two case%
\begin{eqnarray}
\mathcal{A} &=&\int \prod_{i=1}^{4}dz_{i}<e^{ik_{1}X(z_{1})}\epsilon
_{2\mu \nu }\cdot \partial X^{\mu }(z_{2})\partial X^{\nu
}(z_{2})e^{ik_{2}X(z_{2})}e^{ik_{3}X(z_{3})}e^{ik_{4}X(z_{4})}>  \label{7} \\
&=&\int
\prod_{i=1}^{4}dz_{i}<e^{ik_{1}X(z_{1})}e^{ik_{2}X(z_{2})+i\epsilon
_{2}^{(1)}\cdot \partial X(z_{2})+i\epsilon _{2}^{(2)}\cdot \partial
X(z_{2})}e^{ik_{3}X(z_{3})}e^{ik_{4}X(z_{4})}>\mid _{\text{multilinear in }%
\epsilon _{2}^{(1)},\epsilon _{2}^{(2)}}  \label{8} \\
&=&\int_{0}^{1}dz(1-z)^{k_{2}\cdot k_{3}}z^{k_{1}\cdot k_{2}}\left[\frac{%
\epsilon _{2}^{(1)}\cdot k_{1}}{z}-\frac{\epsilon _{2}^{(1)}\cdot k_{3}}{1-z}\right]\left[\frac{\epsilon _{2}^{(2)}\cdot k_{1}}{z}-\frac{\epsilon _{2}^{(3)}\cdot
k_{3}}{1-z}\right]  \label{9}
\end{eqnarray}%
where $\epsilon _{3\mu }^{(l)}\epsilon _{3\nu }^{(j)}$ is to be identified
with $\epsilon _{3\mu \nu }.$ Note that the terms proportional to $%
k_{1}^{\mu }k_{1}^{\nu }$ and $k_{1}^{\mu }k_{3}^{\nu }$ have been
considered previously for the calculation of four tachyons and one vector,
three tachyons amplitudes respectively. The only new term is the one
proportional to $k_{3}^{\mu }k_{3}^{\nu }$, which can be expressed as%
\begin{equation}
\mathcal{A}_{4}=\sum_{a=2}^{\infty }\binom{k_{2}\cdot k_{3}-2}{a-2}(-1)^{a-2}%
\frac{2}{(k_{1}+k_{2})^{2}+2(a-1)}\epsilon _{2\mu \nu }k_{3}^{\mu
}k_{3}^{\nu }.  \label{10}
\end{equation}%
The generating function for this term can be seen from Eq.(\ref{8}) as%
\begin{eqnarray}
G_{2} &=&\exp ^{\left\{ -k_{3}\cdot k_{2}[-\ln (1-z)]\right\} }\exp
^{\left\{ -\epsilon _{2}^{(1)}\cdot k_{3}z\frac{d}{dz}[-\ln (1-z)]\right\}
}\exp ^{\left\{ -\epsilon _{2}^{(2)}\cdot k_{3}z\frac{d}{dz}[-\ln
(1-z)]\right\} }\mid _{\text{multilinear in }\epsilon _{2}^{(1)},\epsilon
_{2}^{(2)}}  \label{11} \\
&=&\left(\epsilon _{2}^{(1)}\cdot k_{3}\right)\,\,z\frac{d}{dz}[\ln (1-z)]\exp ^{\left\{ 
\frac{k_{3}\cdot k_{2}}{2}[\ln (1-z)]\right\} }\left(\epsilon _{2}^{(1)}\cdot
k_{3}\right)\,z\frac{d}{dz}[\ln (1-z)]\exp ^{\left\{ \frac{k_{3}\cdot k_{2}}{2}[\ln
(1-z)]\right\} }  \label{12} \\
&=&\sum_{a=2}^{\infty }\binom{k_{2}\cdot k_{3}-2}{a-2}(-1)^{a-2}\epsilon
_{2\mu \nu }k_{3}^{\mu }k_{3}^{\nu }z^{a}.  \label{13}
\end{eqnarray}%
Eq.(\ref{12}) contains product of two derivative terms which again can be traced
back to $\partial X^{\mu }\partial X^{\nu }$ part of the spin two vertex.
After setting $z=1$ in Eq.(\ref{13}) above, one can match with the correct result
in Eq.(\ref{10}). 

The calculation above can be generalized to arbitrary higher spin vertex. We
thus conclude that generically generating function for Stirling number of
the first kind connects BCFW precription with scalar-like recursion relation
to arbitrary high spin level scattering, provided that the corresponding
derivatives in its worldsheet integral expression are included.

\section{Conclusions}

Starting from the familiar $4$-point Veneziano formula we have
demonstrated that the scalar-like recursion relation observed by
Cheung, O'Connell and Wecht in \cite{Cheung:2010vn} and by
Fotopoulos in \cite{Fotopoulos:2010jz} can indeed be understood from
BCFW on-shell recursion relation of string amplitudes. We showed
that explanation to the absence of higher-spin modes was very much
like a similar mechanism observed in BCFW on-shell recursion
relation of gauge theory amplitudes: While in gauge theory Ward
identity guarantees  that  two unphysical degrees of freedom
necessary to make up for the completeness
relation\cite{Feng:2011twa}
\bea g_{\mu\nu}= \epsilon_\mu^+ \epsilon_\nu^- +\epsilon_\mu^-
\epsilon_\nu^+ +\epsilon_\mu^L \epsilon_\nu^T+\epsilon_\mu^T
\epsilon_\nu^L ~~~\label{vector2}\eea
decouple, in bosonic string amplitude the  No-Ghost Theorem does the
same thing to decouple  necessary unphysical degrees of freedom that
make up for the whole Fock space completeness relation, which makes
the translation between covariant and scalar-behaved on-shell
relations of string amplitudes. The freedom to translate on-shell
recursion relation between Fock state and physical state is
especially of practical interests since writing down polarization
tensors for generic physical high-spin modes can be quite
complicated in string theory context.

Although our method can be used to calculate string scattering
amplitudes using the on-shell recursion relation, it may be not the
best way to do so. However, it could provide another point of view
to discuss some analytic properties of string theory along, for
example, the work of  Benincasa and Cachazo\cite{Benincasa:2007xk},
and the work of Fotopoulous and Tsulaia\cite{Fotopoulos:2010ay},
based on consistency using different BCFW-deformations to calculate
amplitudes. It can also be used to discuss possible loop amplitudes
using unitarity cut method \cite{Uni}.

%%%%%%%%%%%%%%%%%%%%%%%%%%%%%%%%%%%%
\subsection*{Acknowledgements}
%%%%%%%%%%%%%%%%%%%%%%%%%%%%%%%%%%

We would like to thank R. Boels, F. Cachazo, D. Skinner for valuable
 discussions.
B.F would like to thank the hospitality of Perimeter Institute where
this work was presented.  This work is supported, in part, by fund
from Qiu-Shi and Chinese NSF funding under contract No.11031005,
No.11135006, No. 11125523. CF is supported by 
National Science Council, 50 billions project
of Ministry of Education and National Center for Theoretical Science, Taiwan,
Republic of China. We would also like to acknowledge the support 
of S.T. Yau center of NCTU.
%National Center for 
%Theoretical Sciences (North) and National Science
%Council, Taiwan, 

%, under contract NSC
%101-2811-M-009-047.

%%%%%%%%%%%%%%%%%%%%%%%%%%%%%
%%%%%%%%%%%%%%%%%%%%%%%%%%%%%
%%%%%%%%%%%%%%%%%%%%%%%%%%%%%
%%%%%%%%%%%%%%%%%%%%%%%%%%%%%
%%%%%%%%%%%%%%%%%%%%%%%%%%%%%
%%%%%%%%%%%%%%%%%%%%%%%%%%%%%
%%%%%%%%%%%%%%%%%%%%%%%%%%%%%
\appendix
%%%%%%%%%%%%%%%%%%%%%%%%%%%
%%%%%%%%%%%%%%%%%%%%%%%%%%%%%
%%%%%%%%%%%%%%%%%%%%%%%%%%%%%
%%%%%%%%%%%%%%%%%%%%%%%%%%%%%
%%%%%%%%%%%%%%%%%%%%%%%%%%%%%

%%%%%%%%%%%%%%%%%%%%%%
\section{Mathematical identity}
\label{app-mathid}
%%%%%%%%%%%%%%%%%%%%%%

{\bf Stirling Number of the first kind:} The Stirling numbers of the
first kind is defined from the generation function
\bea (x)_n\equiv  x(x-1)...(x-n+1)=\sum_{k=0}^n s(n,k)
x^k~~~\label{Stirling-1-def}\eea
where $(x)_n$ is the {\bf Pochhammer symbol} for the falling
factorial and when $n=0$, $(x)_0\equiv 1$. Using this, we can see
that $s(0,0)=1$ but $s(n,0)=0$ if $n\neq 0$.

The signed Stirling numbers of the first kind  are defined such that
the number of permutations of $n$ elements which contain exactly $m$
permutation cycles is the nonnegative number
\bea |s(n,m)|=(-)^{n-m} s(n,m)= n!\sum_{\{N_t\}}
\prod_{t=1}^{\infty} {1\over N_t! t^{N_t}},~~~~\sum t N_t=n,~~m=\sum
N_t ~~~\label{Stirling-1-exp}\eea

There are other ways to see above identities. Considering following
Taylor expansion
\begin{equation}
I_1= (1-z)^{X}=\sum_{a=0}^{\infty}\left(\begin{array}{c}
X\\
a
\end{array}\right)(-)^{a}z^{a}
\end{equation}
which can be expanded by following alternative way
\begin{align}
\mbox{exp} \big [ \,X\,\mbox{ln}(1-z) \big ]&=\,\mbox{exp} \Bigg [ \,(-X)\,\Big (\,z+\frac{1}{2}z^{2}+\frac{1}{3}z^{3}+ \cdots+\frac{1}{n}z^{n}+\cdots \Big ) \Bigg ]\nonumber\\
	   &=\sum\limits_{N=0}^{\infty} \sum\limits_{J=0}^{N} \frac{\big |s(N,J) \big |}{N!}\,(-X)^{J}\,z^{N}\nonumber\\
	   &=\sum\limits_{N=0}^{\infty} \sum\limits_{J=0}^{N} \frac{s(N,J)}{N!}\,(-)^{N}\,X^{J}\,z^{N} ~~~\label{gen-funct-app}
\end{align}
Comparing these two expansions we can refer \eref{Stirling-1-def}
and \eref{Stirling-1-exp}.
%%%%%%%%%%%%%%%%%%%%%%%%%%%%%%%%%%%%%%%%%%%%%
%%%%%%%%%%%%%%%%%%%%%%%%%%%%%%%%%%%%%%%%%%%%%
%%%%%%%%%%%%%%%%%%%%%%%%%%%%%%%%%%%%%%%%%%%%%
%%%%%%%%%%%%%%%%%%%%%%%%%%%%%%%%%%%%%%%%%%%%%
%%%%%%%%%%%%%%%%%%%%%%%%%%%%%%%%%%%%%%%%%%%%%
%%%%%%%%%%%%%%%%%%%%%%%%%%%%%%%%%%%%%%%%%%%%%
%%%%%%%%%%%%%%%%%%%%%%%%%%%%%%%%%%%%%%%%%%%%%
\section{Decoupling of Ghosts in string amplitude} \label{ddf-ghost}
%%%%%%%%%%%%%%%%%%%%%%%%%%%%%%%%%%%%%%%%

The content in this section can be found in \cite{String}. In
bosonic string theory, physical states are required to satisfy
Virasoro constraints $(L_{0}-1)\left|\phi\right\rangle =0$ and
$L_{m>0}\left|\phi\right\rangle =0$. As we have seen in section
\ref{section-int-pole}, the first of these two types of constraints
was implemented as on-shell condition \eref{mass-level} so that it
is satisfied by intermediate states that appear in BCFW recursion
relation. In this appendix we prove that ghosts decouples from BCFW
recursion relation. As a consequence we are allowed to introduce
freely the physical states, for which the remaining Virasoro
constraint $L_{m>0}\left|\phi\right\rangle =0$  applies, or generic
Fock states as intermediate states in the recursion relation. For
the purpose of argument needed in this proof we first divide Fock
space into three subspaces according to DDF construction.

%%%%%%%%%%%%%%%%
\subsection{ DDF states}
%%%%%%%%%%%%%%%%%%%%%%%%%%
A standard DDF state is defined by acting a string of
transverse $A_{-n}^{i} $ operators on tachyonic vacuum
\bea \ket{f}= A^{i_1}_{-n_1}
A^{i_2}_{-n_2}...A^{i_m}_{-n_m}\ket{0;p_0}~,~~~\label{DDF-state}\eea
where DDF operator $A_{n}^{i} $ is prescribed as the
 Fourier zero mode of  vector vertex operator $V_j(nk_0,\tau)=\dot{X}^j(\tau) e^{in X^+(\tau)}$,
\bea A_n^i={1\over 2\pi}\int_0^{2\pi} {\dot X}^i(\tau) e^{in
X^+(\tau)} d\tau,~~~i=1,...,D-2~,~~~\label{DDF-ope}\eea
and $p_0=(p_0^+, p_0^-, p_0^i)=(1,-1,0)$.
It is easy to show that $L_{m>0}\ket{f}=0$ since $L_m$ commutates
with all $A^i_{-n}$ while $L_{m>0}\ket{0;p_0}=0$. For $L_0$, using
that $L_0\ket{0;p_0^2}=\a' p_0^2=1$ we get $(L_0-1)\ket{0;p_0}=0$.
The DDF states thus defined are positive definite, as can be easily
checked using the commutation relation $[A_m^i, A_n^j]= m\delta_{ij}
\delta_{m+n}$. We shall denote in the following a generic DDF state as
$\left|f\right\rangle$. Note however, that in the standard construction
these DDF states are automatically on the $N$-mass-shell,
\bea \WH p \ket{f}= (p_0+k_0\sum n_i) \ket{f}~~~\label{DDF-mom}\eea
so that $(p_0+N k_0)^2= p_0^2+ 2N=2+2N$, where we introduced
$k_0=(k_0^+,k_0^-, k_0^i)=(0,-1,0)$, and here $N=\sum n_{i}$.
In order to describe Fock states in DDF language, where
center-of-mass momentum $k^{\mu}$  and mode number $N$  are
considered independent, let us define generalized off-shell
DDF-like state, starting again from tachyonic vacuum but
with momentum $q+N\, k_{0}$,
\bea \ket{f}_{off-shell}= A^{i_1}_{-n_1}
A^{i_2}_{-n_2}...A^{i_m}_{-n_m}\ket{0;q+N\,k_{0}}.~~~\label{DDF-like-M-1}\eea
Note that we shift ground state momentum by equal and
opposite of the amount that is going to be
shifted by DDF operators so that subsequent operations
 produces an off-shell state with arbitrary momentum $q$
and mode eigenvalue $N=\sum_{i}\, n_{i}$. In addition to
DDF operators we introduce operators $K_{m}$, defined as
\bea K_m=k_0\cdot \a_m=-\a_{m}^+~~~~\label{K-ope} \eea
and consider states constructed by operating
a string of Virasoro generator $L_{-n}$ and $K_{-m}$ on
DDF-like state $\ket{f}_{off-shell}$ carrying off-shell momentum
$q$ in the following order
\bea \ket{\{\lambda, \mu\}, f} = L_{-1}^{\la_1} L_{-2}^{\la_2}...
L_{-n}^{\la_n} K_{-1}^{\mu_1}...
K_{-m}^{\mu_m}\ket{f}_{off-shell}~.~~\label{LK-DDF}\eea
The set of states $\left|\left\{ \lambda,\mu\right\} ,f\right\rangle$
with $\sum r\lambda_{r}+\sum s\mu_{s}+\sum n_{i}=N$  are linearly
independent and constitutes a basis that spans level-$N$  subspace at
fixed momentum $q$. In the following discussions for convenience
we drop the lower script that distinguishes DDF state $\ket{f}$ and
 DDF-like state $\ket{f}_{off-shell}$, while it is understood that
the center-of-mass momentum is considered as a independent
degree of freedom, on-shell or not, whenever
a DDF basis is referred to.

%%%%%%%%%%%%%%%%%%%
\subsection{Decoupling of ghosts in string amplitude}
%%%%%%%%%%%%%%%%%%%%
States \eref{LK-DDF} can be divided into two types. The first type
is with $L_{-n}$ in front, so it is spurious state
$\ket{s}$. %\footnote{I need to emphasize that here the spurious state
%means only that it is orthogonal to physical states and in general
%$(L_0-1)\ket{s}\neq 0$. I have checked the Polchinski's book, it
%seems his notation of spurious states does not require on-shell
%condition.}.
The second one is without $L_{-n}$ and we denote it as
$\ket{k}$. Thus any state in the Fock space can be uniquely
decomposed as
\bea  \ket{\phi}=\ket{s}+\ket{k}\eea
where $\ket{s}$ is the {\sl spurious} state  and $\ket{k}$ is the
form in \eref{LK-DDF} without any $L_{-n}$ in front of the
expression. Since $\ket{s},\ket{k}$ are linear independently, if
$\ket{\phi}$ is the eigenstate of $L_0$, so are $\ket{s},\ket{k}$.
This means that if
\bea (L_0-1)\ket{\phi}=0,~~~\Longrightarrow
(L_0-1)\ket{s}=(L_0-1)\ket{k}=0 \eea
Next we show that  if the state $\ket{\phi}$ is physical state, the
decomposed states $\ket{s}$ and $\ket{k}$ are also physical states.

Because $\ket{s}$ is spurious  and physical when $\ket{\phi}$
is physical, we have $\Spaa{s|s}=\Spaa{s|k}=0$, so
$\Spaa{\phi|\phi}=\Spaa{k|k}$. We can decompose $\ket{k}=\ket{f}+
\ket{\W k}$ where $\ket{f}$ is DDF state and $\ket{k}$ is the form
of \eref{LK-DDF} without string of $L$ but at least one of $K_{-m}$.
By the property of $K_{-m}$, it is easy to shown that $\Spaa{\W k|\W
k}=\Spaa{\W k|f}=0$, so finally we have
$\Spaa{\phi|\phi}=\Spaa{k|k}=\Spaa{f|f}$. This is the familiar result
known as  the ``No-ghost Theorem'' for  string {\it amplitude},
which can also be characterized as the absence of negative norm
among general physical state $\ket{\phi}$.

In fact, there is a stronger statement. Using $[L_m, K_n]-n K_{m+n}$
and $L_{m>0}\ket{f}=0$, it can show that  if $\ket{k}$ is physical,
then $\ket{\W k}=0$ in the expansion of $\ket{k}=\ket{f}+\ket{\W
k}$. Thus we see that the general physical state $\ket{\phi}$ can be
written
\bea \ket{\phi}=\ket{f}+\ket{s} \eea
where $\ket{f}$ is a DDF state and $\ket{s}$ is a spurious physical
state. The appearance of spurious physical state $\ket{s}$, i.e.,
the transformation $\ket{f}\to \ket{f}+\ket{s}$ is the
string-theoretic analog of a gauge transformation.

%%%%%%%%%%%%%%%%%%%%%%%%%
\subsection{Decoupling of ghosts in BCFW on-shell recursion relation}
\label{no-ghost}
%%%%%%%%%%%%%%%%%%%%%%%%%%

In section \ref{sec-n-pt} we saw that pole structure in a bosonic
string amplitude is manifest when expressed in algebraic form
\bea
A_{M}=\left\langle \phi_{1}\right|V_{2}\Delta V_{3}\dots V_{i}\Delta V_{i+1}\dots V_{M-1}\Delta\left|\phi_{M}\right\rangle .\eea
Residue at the $(i-1)$-th pole at mass level $N$  is therefore given
by the sum of products
\bea
\sum_{\begin{array}{c}
level-N\\
states
\end{array}}\left\langle \phi_{1}(k_{1}+z_{i,N}\,
q)\right|V_{2}\Delta V_{3}\dots V_{i}\left|\{N_{\mu,m}\},
\hat{p}\right\rangle \left\langle \{N_{\mu,m}\},
\hat{p}\right|V_{i+1}\dots \Delta V_{M-1}\left|\phi_{M}
(k_{M}-z_{i,N}\, q)\right\rangle ,~~~\label{residue-sum-product}
\eea
where the above sum is  taken only over intermediate Fock states
that happen to be on the level-$N$  mass-shell. Note that in BCFW
recursion relation the mode eigenvalues $\{N_{\mu,m}\}$  and
center-of-mass momentum $\hat{p}$  of intermediate states were
originally considered as independent. {\it It is because $\{N_{\mu,m}\}$
and $\hat{p}$  assume the values $\sum N_{\mu,m}=N$  and
$\frac{1}{2}\hat{p}^{2}(z)+N-1=0$  that a pole was created at $z=z_{i,N}$
 in the first place}, so that at pole the mass-shell condition
 is automatically satisfied.

Consider the state
\bea \ket{\phi_{R}}= V_{i+1} \Delta V_{i+2}  ....\Delta
V_{M-1}\ket{\phi_M}\eea
that appears on the right side of equation
(\ref{residue-sum-product}). Since we are only interested in its
product with on-shell states, let us operate on it a projection
operator $P_{1}$. For the purpose of proving decoupling of
ghosts, first we would like to show that
\bea  L_{m>0} P_1\ket{\phi_{R}}=0,\eea
where we defined $P_k$ as a projection operator  which projects
states to subspace with $L_0=k$.
Using $[L_0,L_m]= -m L_{m}$, we find $L_0 L_m P_1\ket{\a}=(1-m) L_m
P_1\ket{\a}$, so $L_m P_1=P_{1-m} L_m P_1=P_{1-m} L_m$, thus we need
to prove
\bea P_{1-m} L_m\ket{\phi_{R}}=0,~~~~m>0 \eea
Using $P_{1-m} (-L_0-m+1)=0$, we get
\bea P_{1-m} (L_m-L_0-m+1)\ket{\phi_{R}}=0,~~~~m>0 \eea
Finally we arrive at the  identity
\bea (L_m-L_0-m+1)V_N \Delta V_{N+1}  ....\Delta
V_{M-1}\ket{\phi_M}=0,~~~~m>0\eea
Note that a vertex $V$ has conformal dimension one, therefore satisfies
\bea [L_m, V(k,z)] =\left( z^{m+1} {d\over dz}+ m
z^m\right) V(k,z).~~~\label{L-V-rel} \eea
Now using the \eref{L-V-rel} and set $z=1$ (since we have $\tau=0$
which is crucial) we have
\bea [L_m-L_0, V]=m
V,~~~or~~(L_m-L_0-m+1)V=V(L_m-L_0+1)~~~~\label{LV-1}\eea
where ${d\over dz}$ has  been canceled. Using  Virasoro algebra
it is straightforward to show that
\bea (L_m-L_0+1){1\over L_0-1}={1\over L_0+m-1}
(L_m-L_0-m+1)~~~~\label{LV-2}\eea
Thus \eref{LV-1} and \eref{LV-2} give
\bea (L_m-L_0-m+1)V{1\over L_0-1}=V{1\over L_0+m-1}
(L_m-L_0-m+1)\eea
so $(L_m-L_0-m+1)$ can be pushed step by step all the way to the
right until it meets $\ket{\phi_M}$, and we obtain
 $(L_m-L_0+1)\ket{\phi_M}=0$ because
$\ket{\phi_M}$ is physical. From the argument above we
see that when on-shell, $\ket{\phi_M}$ satisfies Virasoro
constraints and is therefore a physical state. It is
straightforward to see that the same
argument applies to state $\ket{\phi_L}= V_{i} \Delta V_{i-1}  ....\Delta
V_{2}\ket{\phi_{1}}$.

~\\

{\bf Proof:}
Having done all the preparations we are now finally ready to
derive our proof. We note that in the algebraic expression
(\ref{residue-sum-product}) for residue at mass level $N$,
the summation of outer products of Fock states
$\left|\{N_{\mu,m}\},\hat{p}\right\rangle
\mathcal{T}_{\{N_{\mu,m}\}}\left\langle \{N_{\mu,m}\},\hat{p}\right|$
over level-$N$  subspace works as a projection operator that
maps $\left|\phi_{R}\right\rangle$   and $\left|\phi_{L}\right\rangle$
into the level-$N$  subspace, so that if we decompose in this
sector $\left|\phi_{R}\right\rangle$   and $\left|\phi_{L}\right\rangle$
according to DDF basis into $\left|s\right\rangle +\left|\tilde{k}\right\rangle
+\left|f\right\rangle$, the residue (\ref{residue-sum-product}) reads
\bea
\begin{array}{c}
\sum_{\begin{array}{c}
level-N\\
states
\end{array}}\left\langle \phi_{L}|\{N_{\mu,m}\},\hat{p}\right\rangle \mathcal{T}_{\{N_{\mu,m}\}}\left\langle \{N_{\mu,m}\},\hat{p}|\phi_{R}\right\rangle =\left\langle s_{L}+\tilde{k}_{L}+f_{L}|s_{R}+\tilde{k}_{R}+f_{R}\right\rangle \\
=\left\langle f_{L}|f_{R}\right\rangle. ~~~\label{ghost-dec-bcfw1}
\end{array}
\eea
As argued in the decoupling of ghosts in amplitudes, spurious
state $\left|s\right\rangle$   drop out from (\ref{ghost-dec-bcfw1})
because both $\left|\phi_{R}\right\rangle$   and $\left|\phi_{L}\right\rangle$
are physical, and we remove subsequently $\left|\tilde{k}\right\rangle$
states since
$\left\langle \tilde{k}|\tilde{k}\right\rangle
=\left\langle \tilde{k}|f\right\rangle =0 $.

Inserting complete states again, but this time in DDF basis,
into the product $\left\langle f_{L}|f_{R}\right\rangle$,
\bea
\left\langle f_{L}|f_{R}\right\rangle
& = & \sum_i\Spaa{f_L|s_i+k_i+f_i}
\Spaa{s_i+k_i+f_i|f_R} \nn
& = & \sum_i\Spaa{f_L|f_i} \Spaa{f_i|f_R}
 =  \sum_i\Spaa{f_L+s_L|f_i}
\Spaa{f_i|f_R+s_R} \nn
& = & \sum_i\Spaa{\phi_L|f_i} \Spaa{f_i|\phi_R}
\eea
and we see that spurious and $\left|\tilde{k}\right\rangle$
intermediate states drop out for the same reason, thus
summing over the whole intermediate Fock space is equivalent
to summing over the physical subspace.

\end{document}